\pdfoutput=1
\documentclass[
paper=letter,%
pagesize,%
numbers=noendperiod,%
captions=nooneline,%
abstracton%
]{scrartcl}

\usepackage[T1]{fontenc}%
\usepackage{lmodern}%
\usepackage[american]{babel}%
\usepackage{microtype}%

\usepackage[
hyperref,%
table%
]{xcolor}%

\usepackage[nouppercase]{scrpage2}%

\usepackage{eso-pic}%
\usepackage{rotating}%
\usepackage{amsmath}%
\usepackage{mathtools}%
\usepackage{amssymb}%
\usepackage{amsthm}%
\usepackage{etoolbox}%
\usepackage{bm}%
\usepackage{bbm}%
\usepackage{enumitem}%
\usepackage{graphicx}%
\usepackage{grffile}%
\usepackage{tikz}%
\usepackage{wrapfig}%
\usepackage{tabularx}%
\usepackage{siunitx}%
\usepackage{booktabs}%
\usepackage{multirow}%
\usepackage{vruler}%
\usepackage{fancyvrb}%
\usepackage{listings}%
\usepackage{csquotes}%
\usepackage[
backend=bibtex,%
style=authoryear,%
dashed=false,%
uniquename=init,%
firstinits=true,%
hyperref=true,%
useprefix=true,%
maxcitenames=2,%
maxbibnames=6,%
date=iso8601,%
urldate=iso8601%
]{biblatex}%
\usepackage[
hypertexnames=false,%
setpagesize=false,%
pdfborder={0 0 0},%
pdfstartview=Fit,%
bookmarksopen=true,%
bookmarksnumbered=true%
]{hyperref}%
\xdefinecolor{lightgray}{RGB}{247, 247, 247}%
\xdefinecolor{semilightgray}{RGB}{240, 240, 240}%
\xdefinecolor{middlegray}{RGB}{127, 127, 127}%
\xdefinecolor{blue}{RGB}{58, 95, 205}%
\xdefinecolor{deepskyblue}{RGB}{0, 154, 205}%
\xdefinecolor{chocolate}{RGB}{205, 102, 29}%

\setkomafont{pageheadfoot}{\normalfont\normalcolor\sffamily}%
\setkomafont{pagenumber}{\normalfont\normalcolor\sffamily}%
\automark{section}%
\setkomafont{captionlabel}{\normalfont\normalcolor\sffamily\bfseries}%

\setlist{%
  align=left,%
  labelsep=*,%
  leftmargin=*,%
  topsep=1mm,%
  itemsep=0mm%
}
\newcommand*{\mysquare}{\rule[0.18em]{0.36em}{0.36em}}
\newcommand*{\mytriangle}{\raisebox{0.12em}{\resizebox{0.48em}{0.48em}{$\blacktriangleright$}}}
\newcommand*{\mybar}{\rule[0.32em]{0.62em}{0.08em}}
\newcommand*{\mydot}{\raisebox{0.14em}{\resizebox{0.44em}{!}{$\bullet$}}}
\setlist[itemize,1]{label={\mysquare\ }}%
\setlist[itemize,2]{label={\mytriangle\ }}%
\setlist[itemize,3]{label={\mybar\ }}%
\setlist[itemize,4]{label={\mydot\ }}%
\setlist[enumerate,1]{label=\arabic*)}%
\setlist[enumerate,2]{label=\arabic{enumi}.\arabic*)}%
\setlist[enumerate,3]{label=\arabic{enumi}.\arabic{enumii}.\arabic*)}%

\makeatletter
\newcommand\myisodate{\number\year-\ifcase\month\or 01\or 02\or 03\or 04\or 05\or 06\or 07\or 08\or 09\or 10\or 11\or 12\fi-\ifcase\day\or 01\or 02\or 03\or 04\or 05\or 06\or 07\or 08\or 09\or 10\or 11\or 12\or 13\or 14\or 15\or 16\or 17\or 18\or 19\or 20\or 21\or 22\or 23\or 24\or 25\or 26\or 27\or 28\or 29\or 30\or 31\fi}%
\makeatother
\newcommand*{\abstractnoindent}{}%
\let\abstractnoindent\abstract
\renewcommand*{\abstract}{\let\quotation\quote\let\endquotation\endquote
  \abstractnoindent}
\deffootnote[1em]{1em}{1em}{\textsuperscript{\thefootnotemark}}%
\lstdefinestyle{input}{
  backgroundcolor=\color{semilightgray},%
  commentstyle=\itshape\color{chocolate},%
  keywordstyle=\color{blue},%
  stringstyle=\color{blue},%
  numbers=left,%
  numbersep=4.8pt,%
  numberstyle=\color{darkgray!80}\tiny%
}
\lstdefinestyle{output}{
  backgroundcolor=\color{lightgray}%
}

\lstdefinestyle{Lstyle}{
  language=[LaTeX]TeX,%
  texcs={},%
  otherkeywords={}%
}

\lstnewenvironment{Linput}[1][]{%
  \lstset{style=input, style=Lstyle}
  #1%
}{\vspace{-0.25\baselineskip}}%

\lstnewenvironment{Loutput}[1][]{%
  \lstset{style=output, style=Lstyle}
  #1%
}{\vspace{-0.25\baselineskip}}%

\lstdefinestyle{Rstyle}{
  language=R,%
  keywords={if, else, repeat, while, function, for, in, next, break},%
  otherkeywords={}%
}

\lstnewenvironment{Rinput}[1][]{%
  \lstset{style=input, style=Rstyle}
  #1%
}{\vspace{-0.25\baselineskip}}%

\lstnewenvironment{Routput}[1][]{%
  \lstset{style=output, style=Rstyle}
  #1%
}{\vspace{-0.25\baselineskip}}%

\DeclareNameAlias{sortname}{last-first}%
\DefineBibliographyExtras{american}{\DeclareQuotePunctuation{}}%
\renewbibmacro*{volume+number+eid}{%
  \setunit*{\addcomma\space}%
  \printfield{volume}%
  \printfield{number}}
\DeclareFieldFormat*{number}{(#1)}
\DeclareFieldFormat*{title}{#1}%
\DeclareFieldFormat{doi}{%
  \ifhyperref
    {\href{http://dx.doi.org/#1}{\nolinkurl{doi:#1}}}%
    {\nolinkurl{doi:#1}}}%
\renewbibmacro*{in:}{}%
\DeclareFieldFormat{isbn}{ISBN #1}%
\DeclareFieldFormat{pages}{#1}%
\DeclareFieldFormat{url}{\url{#1}}%
\DeclareFieldFormat{urldate}{\mkbibparens{#1}}%
\renewcommand*{\cite}[2][]{\textcite[#1]{#2}}%

\newif\ifstarttheorem
\newtheoremstyle{mythmstyle}%
{0.5em}%
{0.5em}%
{}%
{}%
{\sffamily\bfseries\global\starttheoremtrue}%
{}%
{\newline}%
{\thmname{#1}\ \thmnumber{#2}\ \thmnote{(#3)}}%
\theoremstyle{mythmstyle}%
\newtheorem{definition}{Definition}[section]%

\newtheorem{remark}[definition]{Remark}

\makeatletter
\preto\itemize{%
  \if@inlabel
    \ifstarttheorem
      \mbox{}\par\nobreak\vskip\glueexpr-\parskip-\baselineskip+0.25em\relax\hrule\@height\z@
    \fi%
  \fi%
  \global\starttheoremfalse%
 \def\tempa{proof}%
 \ifx\tempa\mycurrenvir
    \ifstarttheorem
      \mbox{}\par\nobreak\vskip\glueexpr-\parskip-\baselineskip+0.25em\relax\hrule\@height\z@
    \fi%
 \fi%
 \global\starttheoremfalse%
}
\preto\enditemize{\global\starttheoremfalse}
\makeatother

\makeatletter
\preto\enumerate{%
  \if@inlabel
    \ifstarttheorem
      \mbox{}\par\nobreak\vskip\glueexpr-\parskip-\baselineskip+0.25em\relax\hrule\@height\z@
    \fi%
  \fi%
  \global\starttheoremfalse%
 \def\tempa{proof}%
 \ifx\tempa\mycurrenvir
    \ifstarttheorem
      \mbox{}\par\nobreak\vskip\glueexpr-\parskip-\baselineskip+0.25em\relax\hrule\@height\z@
    \fi%
 \fi%
 \global\starttheoremfalse%
}
\preto\endenumerate{\global\starttheoremfalse}
\makeatother

\newcommand{\ou}[3]{%
  \mathrel{%
    \vcenter{\offinterlineskip
      \ialign{##\cr$#1$\cr\noalign{\kern-#3}$#2$\cr}%
    }%
  }%
}
\newcommand*{\omu}[3]{\underset{#3}{\overset{#1}{#2}}}

\newcommand*{\isim}{\omu{\text{\tiny{ind.}}}{\sim}{}}

\newcommand*{\IR}{\mathbbm{R}}

\newcommand*{\U}{\operatorname{U}}

\newcommand*{\ARMA}{\operatorname{ARMA}}
\newcommand*{\GARCH}{\operatorname{GARCH}}

\renewcommand*{\P}{\mathbbm{P}}

\newcommand{\field}[1]{\mathbb{#1}}
\newcommand{\Reals}{\field{R}}

\newcommand*{\R}{\textsf{R}}
\newcommand*{\SP}{S\&P~500}
\newcommand*{\eps}{\varepsilon}

\begin{document}
\thispagestyle{plain}
\begin{center}
  \sffamily
  {\bfseries\LARGE Visualizing Dependence in High-Dimensional Data: An
    Application to S\&P 500 Constituent Data\par}
  \bigskip\smallskip
  {\Large Marius Hofert\footnote{Department of Statistics and Actuarial Science, University of
    Waterloo, 200 University Avenue West, Waterloo, ON, N2L
    3G1,
    \href{mailto:marius.hofert@uwaterloo.ca}{\nolinkurl{marius.hofert@uwaterloo.ca}}. The
    author would like to thank NSERC for financial support for this work through Discovery
    Grant RGPIN-5010-2015.},
    Wayne Oldford\footnote{Department of Statistics and Actuarial Science, University of
    Waterloo, 200 University Avenue West, Waterloo, ON, N2L
    3G1,
    \href{mailto:woldford@uwaterloo.ca}{\nolinkurl{rwoldford@uwaterloo.ca}}.}\par
    \bigskip
    \myisodate\par}
\end{center}
\par\bigskip
\begin{abstract}
  The notion of a zenpath and a zenplot is introduced to search and detect
  dependence in high-dimensional data for model building and statistical
  inference. By using any measure of dependence between two random variables
  (such as correlation, Spearman's rho, Kendall's tau, tail dependence etc.), a
  zenpath can construct paths through pairs of variables in different ways,
  which can then be laid out and displayed by a zenplot. The approach is
  illustrated by investigating tail dependence and model fit in constituent data of the \SP\
  during the financial crisis of 2007--2008. The corresponding Global Industry
  Classification Standard (GICS) sector information is also addressed.

  Zenpaths and zenplots are useful tools for exploring dependence in
  high-dimensional data, for example, from the realm of finance, insurance and quantitative
  risk management. All presented algorithms are implemented using the \R\
  package \texttt{zenplots} and all examples and graphics in the paper can be
  reproduced using the accompanying demo \texttt{SP500}.
\end{abstract}
\minisec{Keywords}
Zenpath, zenplot, detecting dependence, high dimensions, graphical tools.

\minisec{MSC2010}
62-09, 62H99, 65C60%

\section{Introduction}\label{sec:intro}
Motivated by the use of high-dimensional data such as data from several hundred
risk-factor changes in the realm of quantitative risk management, we raise the
following simple question:
\begin{quote}
  How can one detect and visualize dependence in high-dimensional data?
\end{quote}

Detecting and visualizing dependence in high-dimensional data is important for
model building and inference in areas such as finance, insurance and
quantitative risk
management, where the one-period ahead behaviour of a high-dimensional portfolio represented by a
random vector $\bm{X}=(X_1,\dots,X_d)$, $d$ large, is studied; see
\cite[Section~2.2.1]{mcneilfreyembrechts2015}.

The example we consider in this work is that of detecting and visualizing tail
dependence of a portfolio consisting of sign-adjusted log-returns of all
constituents of the \SP, so we have realizations $(X_{t,j})_t$ of $X_j = X_{t,j} = -\log(S_{t,j}/S_{{t-1,j}})$,
$j\in\{1,\dots,d\}$, for $d\approx 500$, %
and $S_{t,j}$ denotes the end-day price of constituent $j$ of the \SP\ at time
$t\in\{1,\dots,T\}$. To each component of this high-dimensional data, we fit an
$\ARMA(1,1)-\GARCH(1,1)$ model, extract the corresponding standardized residuals and
(for illustration) investigate their tail dependence in the so-called
\emph{copula-GARCH} framework (see \cite{patton2006} and
\cite{patton2013}) %
by considering the pseudo-observations of the standardized residuals. We also
address graphical assessment of model fit.

The high dimensions we consider in this paper are of the order of
hundreds.  Dimensionality reduction is often not a practical option when, for
example, components $X_j$, $j\in\{1,\dots,d\}$, are each individually of
interest and need to be tracked (such as for portfolios of life insurance contracts). Computationally, this
is already fairly demanding, requiring, for example, efficient algorithms to
estimate a (Student) $t$ copula in large dimensions.  Providing meaningful
displays of such high dimensional output adds to this challenge.

Our running example will be the \SP\ constituent data.  The source, the
time-series models we fit, and the connection with copulas, are all described in
Section~\ref{sec:data}.  This setup provides the high-dimensional model output
that will then be visualized.  Section~\ref{sec:zenplot} presents the zenplot, a
compact graphical presentation of high-dimensional data arranged in a path of
one- and two-dimensional displays.  When dimensions are high it will often be
necessary to draw attention to, and to display, only the most salient features
of the data.  To this end, Section~\ref{sec:zenpath} introduces the notion of a
zenpath, a tool to provide an interesting path through the pairs of variables.
In Section \ref{sec:model:check}, zenpaths are then applied in the context of model
assessment and model comparison.   Different  measures are used to illustrate
a variety of zenplots that naturally arise in model criticism and selection whenever the
number of dimensions becomes large. All graphs in this paper can be reproduced with the demo
\texttt{SP500} provided in the \R\ package \texttt{zenplots}; see
\cite{zenplots}. In the last section we provide concluding remarks.

\section{S\&P~500 constituent data}\label{sec:data}
We consider time series of all 505 constituents of the \SP\ as of 2015-10-12;
see \url{https://en.wikipedia.org/wiki/List_of_S\%26P_500_companies} and note
that the \SP\ does not necessarily have exactly 500 constituents. However,
our interest lies in the 756 trading days between 2007-01-01 ($t=0$) and
2009-12-31 ($t=T$) which contains the global financial crisis of 2007--2008.
This data was downloaded from a publicly available source
(\url{https://finance.yahoo.com/} on 2016-01-03) and then incorporated into the
\R\ package \texttt{qrmdata}; see \cite{qrmdata}. We assume that the order of
the data is according to their Global Industry Classification Standard (GICS)
information (see the demo \texttt{SP500} of the \R\ package \texttt{zenplots}).
Of course any company joining the \SP\ after 2009-12-31 will not appear in this
period and, for the remainder there can be much data missing.

\begin{figure}[htbp]
  \centering
  \includegraphics[width=0.48\textwidth]{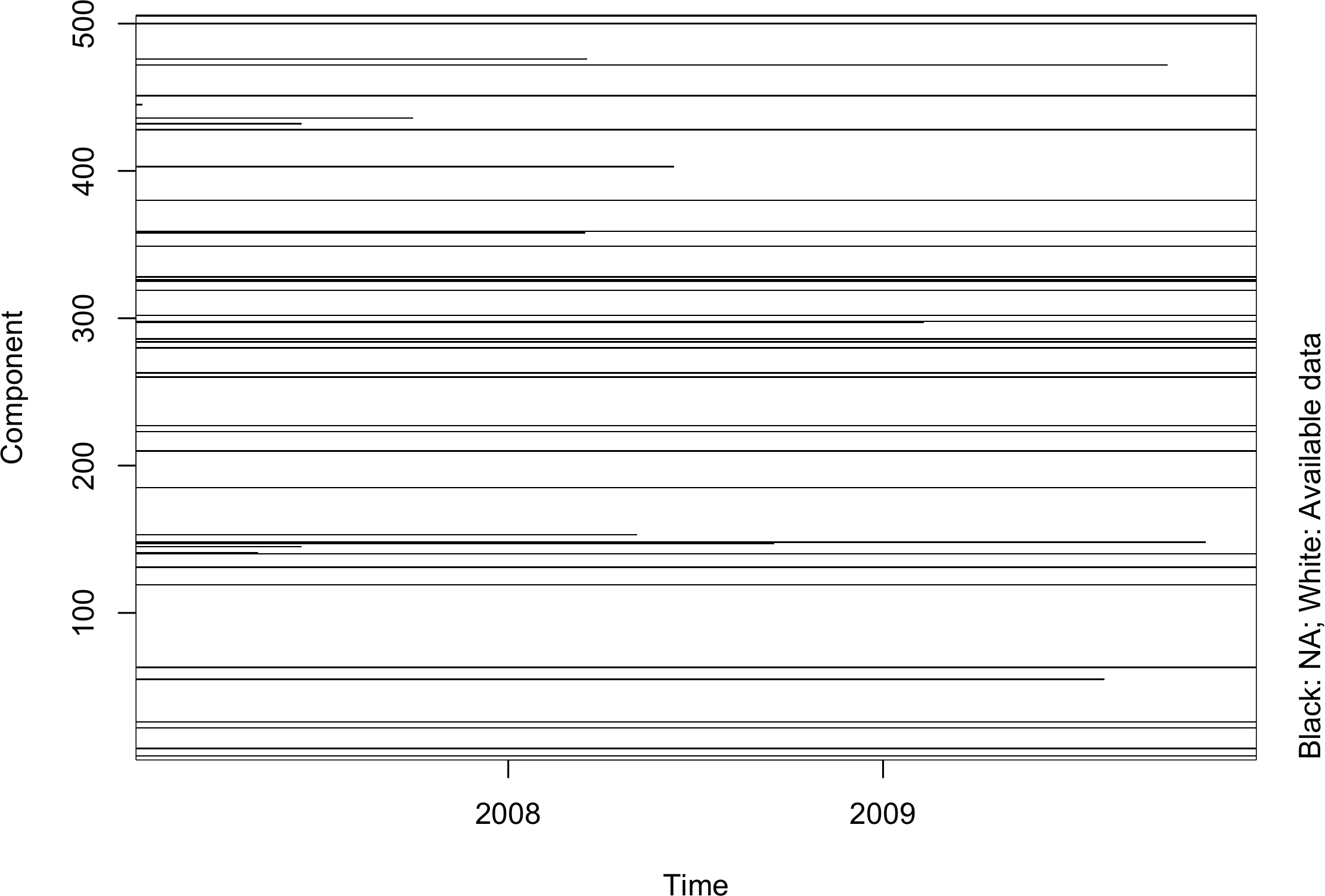}%
  \hfill
  \includegraphics[width=0.48\textwidth]{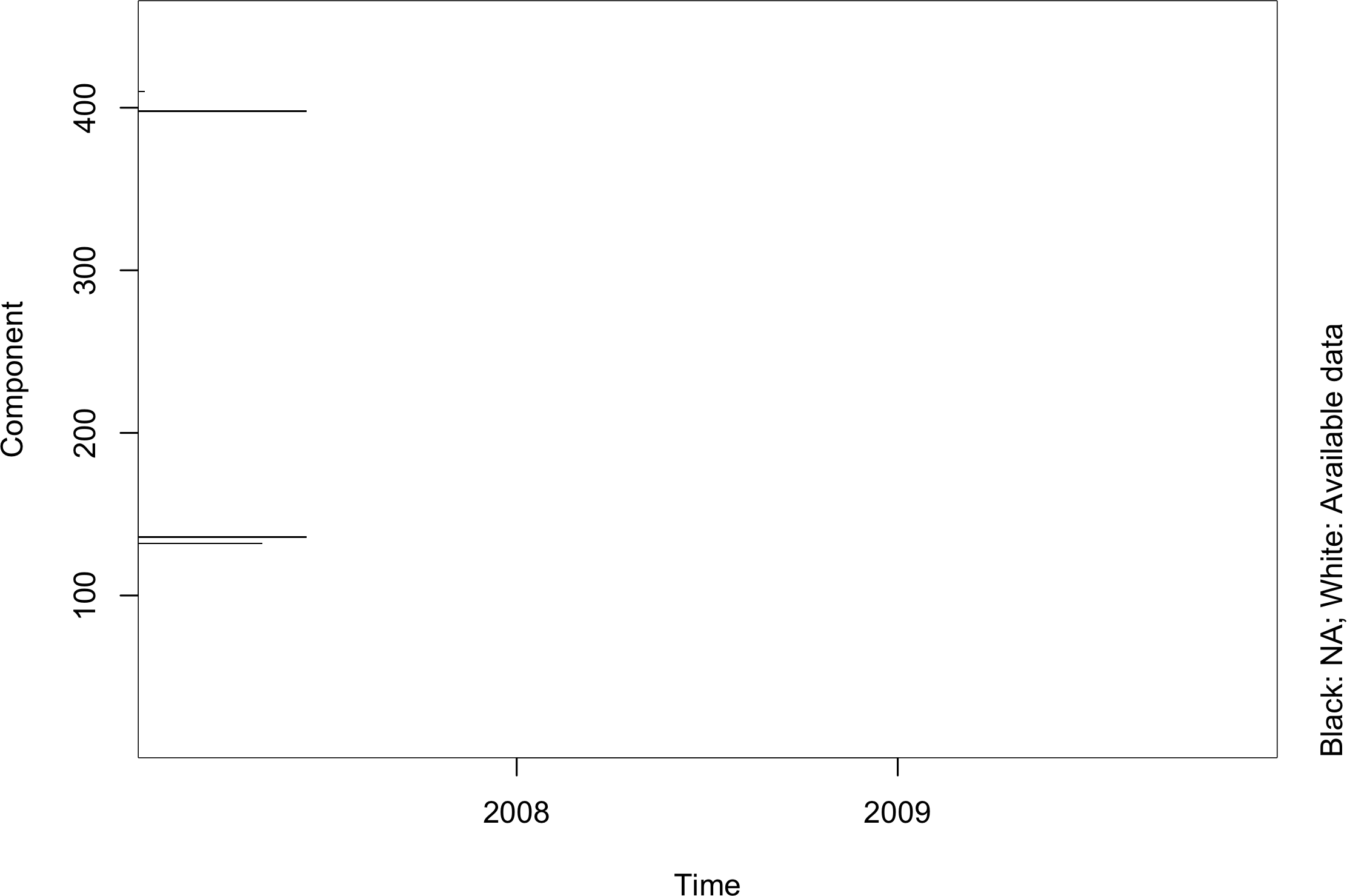}%
  \caption{Missing data for all 505 \SP\ constituents from 2007-01-01 to
    2009-12-31 (left-hand side); days with missing data are marked for each constituent. We work
    with the 465 constituents with maximally 20\% missing data (right-hand side)
    and fill the missing data via \texttt{na.fill(, fill = "extend")}.}
  \label{fig:SP500:NA}
\end{figure}

To assess the extent of the missing data, we plot in
Figure~\ref{fig:SP500:NA} (left-hand side)
the days missing a value for each of the 505 constituents. %
The missing data pattern likely indicates that many of the constituents joined
the \SP\ in this period (or after for those whose lines span the time
period). We are still left with 465 companies if we retain only those having at
least 80\% complete data. As the right-hand side of Figure~\ref{fig:SP500:NA}
shows, only four of these have any data missing (those are \texttt{DAL}
(Delta Air Lines), \texttt{DFS} (Discover Financial Services), \texttt{TEL}
(TE Connectivity Ltd.) and \texttt{TWC} (Time Warner Cable)). We restrict our
analysis to these 465 constituents and fill the missing data of the four
components with \R's \texttt{na.fill(, fill = "extend")}. This function
interpolates linearly between adjacent data where available and, otherwise,
repeats the leftmost or rightmost available data. In the end, we will work
with a complete data set having $T=756$ daily records on $d=465$ dimensions.

\subsection{Modelling the margins}\label{sec:model:margins}
We will first model the 755 negative log-returns for each of the 465 constituents.
Given our time horizon and the financial crisis of 2007--2008, we cannot assume
each constituent series of negative log-returns to be realizations of independent and identically
distributed (iid) random variables.
Instead, we need to incorporate some serial (temporal) dependence for each series.
Because of the large number of marginal time series, we take a broad-brush
approach and use the popular $\ARMA(1,1)-\GARCH(1,1)$ model for each
margin separately (a process also known as de-GARCHing).

These models have the form
\begin{align*}
    X_{t,j}&=\mu_{t,j}+\eps_{t,j}\quad\text{for}\quad\eps_{t,j}=\sigma_{t,j}Z_{t,j},\\
    \mu_{t,j}&=\mu_j+\phi_{j}(X_{t-1,j}-\mu_j)+\theta_{j}(X_{t-1,j}-\mu_{t-1,j}),\\
    \sigma_{t,j}^2&=\alpha_{j0}+\alpha_{j1}(X_{t-1,j}-\mu_{t-1,j})^2+\beta_{j}\sigma_{t-1,j}^2,
\end{align*}
where, for all components $j\in\{1,\dots,d\}$, $\mu_j\in\IR$, $|\phi_j|<1$, $|\theta_j|<1$,
$\alpha_{j0}>0$, $\alpha_{j1}\ge0$, $\beta_{j}\ge0$,
$\alpha_{j1}+\beta_{jk}<1$.

The stochastic component $Z_{t,j}$ for each series $j\in\{1,\dots,d\}$ are the
\emph{innovations}; their empirical counterparts on which our analysis is based
later on are known as \emph{standardized residuals}.  For each series the
innovations are
taken to be iid, centred about zero and having unit variance.  We model the
innovation distribution as a scaled (Student) $t$ distribution; that is, for
each $j\in\{1,\dots,d\}$,
$Z_{t,j}\isim F_{j}(z)=t_{\nu_j}(z \sqrt{\nu_j/(\nu_j-2)})$, where $t_{\nu_j}$
is the distribution function of the standard $t$ distribution with $\nu_j$
degrees of freedom.

Even though each series is fit separately, the fitting itself is non-trivial for
so many components.  We thus use the robust \texttt{fit\_ARMA\_GARCH()} from
\texttt{qrmtools} (developed for this purpose; see \cite{qrmtools}) with
\texttt{solver = "hybrid"} for the underlying fitting procedure
\texttt{ugarchfit()} of the \R\ package \texttt{rugarch} of \cite{rugarch}. The
six warnings which appear can safely be ignored here as they only indicate
issues in finding initial values for the fitting; see the demo for more
details. The estimated standardized residuals
$\widehat{\bm{Z}}_t=(\widehat{Z}_{t,1},\dots,\widehat{Z}_{t,d})$,
$t\in\{1,\dots,T\}$ from this fit will be treated as realizations of $\bm{Z}_t$
in any subsequent analysis. Residual checks are presented in
Section~\ref{sec:model:check:margins:t}.

\subsection{Modelling cross-sectional dependence}
Having modelled the component series marginally, we turn our attention to the
multivariate series of standardized residuals $(\bm{Z}_t)_t$ for iid
$\bm{Z}_t=(Z_{t,1},\dots,Z_{t,d})\sim H$, where $H$ has continuous margins
$F_j$, $j\in\{1,\dots,d\}$, as described before. By Sklar's Theorem, see
\cite{sklar1959}, the joint distribution function $H$ of $\bm{Z}_t$ can be
decomposed as
\begin{align*}
  H(z_1,\dots,z_d)&= C(F_1(z_1),\dots,F_d(z_d)),\quad\bm{z}\in\IR^d \\
                             &= C(u_1, \ldots, u_d),\quad\bm{u}\in [0,1]^d,
\end{align*}
for some copula $C$, where $u_j=F_j(z_j)$, $j\in\{1,\dots,d\}$. The copula $C$
determines the dependence between $Z_{t,1},\dots,Z_{t,d}$.  Since
$u_j = F_j(z_j)$, the copula is itself a distribution function having marginal
$\U(0,1)$ distributions.

In practice, we do not observe data from $C$, but rather from $H$. So to
estimate $C$ we work with
\emph{pseudo-observations} constructed from the available standardized
residuals $\widehat{\bm{Z}}_t=(\widehat{Z}_{t,1},\dots,\widehat{Z}_{t,1})$,
$t\in\{1,\dots,T\}$. Based on the estimated margins $\widehat{F}_j$ of $F_j$,
$j\in\{1,\dots, d\}$, the pseudo-observations of $C$ are computed via
$\widehat{F}_j(\widehat{Z}_{t,j})$, $t\in\{1,\dots,T\}$, $j\in\{1,\dots,d\}$.
One could take the fitted
scaled $t$ distributions mentioned earlier as $\widehat{F}_1,\dots,\widehat{F}_d$. However,  model misspecification  could affect
estimation of $C$, particularly for large dimensions.
Consequently, we prefer to work with the marginal empirical distribution functions $\widehat{F}_{T,j}$  for $j\in\{1,\dots,d\}$ and their pseudo-observations
\begin{align}
  U_{t,j}=\frac{T}{T+1}\widehat{F}_{T,j}(\widehat{Z}_{t,j})=\frac{R_{t,j}}{T+1}\label{pobs}
\end{align}
where $R_{t,j}$ denotes the rank of $\widehat{Z}_{t,j}$ among
$\widehat{Z}_{1,j},\dots,\widehat{Z}_{T,j}$ (the scaling factor $\frac{T}{T+1}$ avoids
having any estimated value of a distribution function be exactly one).
Moreover,  \cite{genestsegers2010} have shown that estimators based on these
pseudo-observations can be asymptotically more efficient (even
when the marginal distributions are known).

The observed joint distribution of the pseudo-observations can then be used to
provide insight on the copula function $C$ and hence the underlying stationary
cross-sectional dependence.

\section{Visualizing dependence in high dimensions with zenplots}\label{sec:zenplot}
\begin{figure}[htbp]
  \centering
  \includegraphics[width=\textwidth]{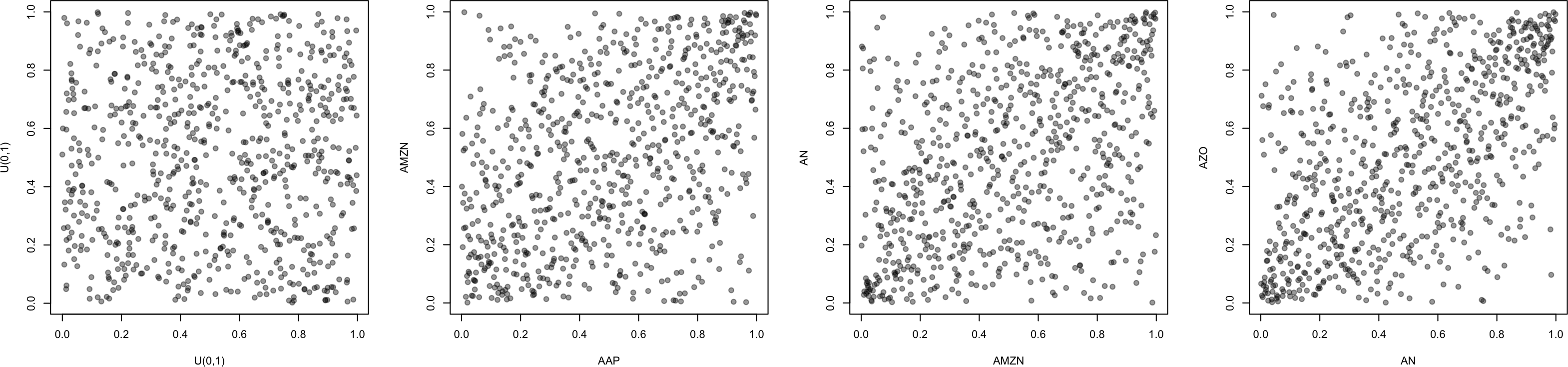}
   \begin{tabular}{p{0.2285\textwidth}p{0.2285\textwidth}p{0.2285\textwidth}p{0.2285\textwidth}}%
   \scriptsize{(a) $\U(0,1) \times \U(0,1)$}  &
   \scriptsize{(b)  $({U}_{t,1}, {U}_{t,2})$ }  %
   &
    \scriptsize{(c)  $({U}_{t,2}, {U}_{t,3})$ } %
    &
   \scriptsize{(d)  $({U}_{t,3}, {U}_{t,4})$ }%
\\
    &
      \scriptsize{~~~~~ = (\texttt{AAP}, \texttt{AMZN}) }%
     &
      \scriptsize{~~~~~ = (\texttt{AMZN}, \texttt{AN}) } %
    &
      \scriptsize{~~~~~ = (\texttt{AN}, \texttt{AZO}) }%
   \end{tabular}
 \caption{Scatterplots of (a) independent $\U(0,1)$ random variables
    and (b, c, d) the pseudo-observation pairs $({U}_{t, j}, {U}_{t, j+1})$,
    $j\in\{1,2,3\}$.
    Ticker symbol abbreviations: \texttt{AAP} = Advanced Auto Parts, \texttt{AMZN} = Amazon.com
    Inc., \texttt{AN} = AutoNation Inc., and \texttt{AZO} = AutoZone
    Inc.}
  \label{fig:SP500:plots:pobs}
\end{figure}

A scatterplot of the pseudo-observations $U_{t,j}$, $t\in\{1,\dots,T\}$,
$j\in\{1,\dots,d\}$ can reveal a lot of information about the dependence structure between any
pair of variates and can give some sense of what features the underlying unknown
copula model has. To simplify the demonstration of this based on our \SP\ data,
consider all columns (or components) sorted according to their GICS sectors in
alphabetical order (together with the number of constituents of that sector):
``Consumer discretionary'' (78), ``Consumer staples'' (33), ``Energy'' (36),
``Financials'' (85), ``Health care'' (51), ``Industrials'' (63), ``Information
technology'' (60), ``Materials'' (25), ``Telecommunications services'' (5), and
``Utilities'' (29).  Within each sector, the original order of the components
remains untouched.  From left to right, Figure~\ref{fig:SP500:plots:pobs}
shows a scatterplot of independent $\U(0,1)^2$ observations followed
by plots of the pseudo-observations for a few pairs of the \SP\ constituents'
standardized residuals. If the standardized residuals from a pair of stocks are statistically independent
from one another, then this plot should be indistinguishable from points
uniformly distributed over the unit square as, for example, they are in
the left-most scatterplot of Figure~\ref{fig:SP500:plots:pobs}.

Roughly speaking, the less the pseudo-observations look like independent
uniforms the stronger is their dependence. For example, each of the three
right-most plots in Figure~\ref{fig:SP500:plots:pobs} departs some from a
uniform, though not dramatically. Each is a little sparser in the top left and
bottom right corners and is a little denser in the bottom left and top right
corners. Though somewhat weak, this shows a positive dependence
between the standardized residuals of these pairs of stocks over this
period in the sense that they tend to be jointly large or small.
The strongest of the three appears to be the right-most plot in
Figure~\ref{fig:SP500:plots:pobs} -- as one might expect between the
standardized residuals of AutoNation Inc.\ (\texttt{AN}) and AutoZone Inc.\
(\texttt{AZO}).

\begin{figure}[htbp]
  \centering
  \includegraphics[width=\textwidth]{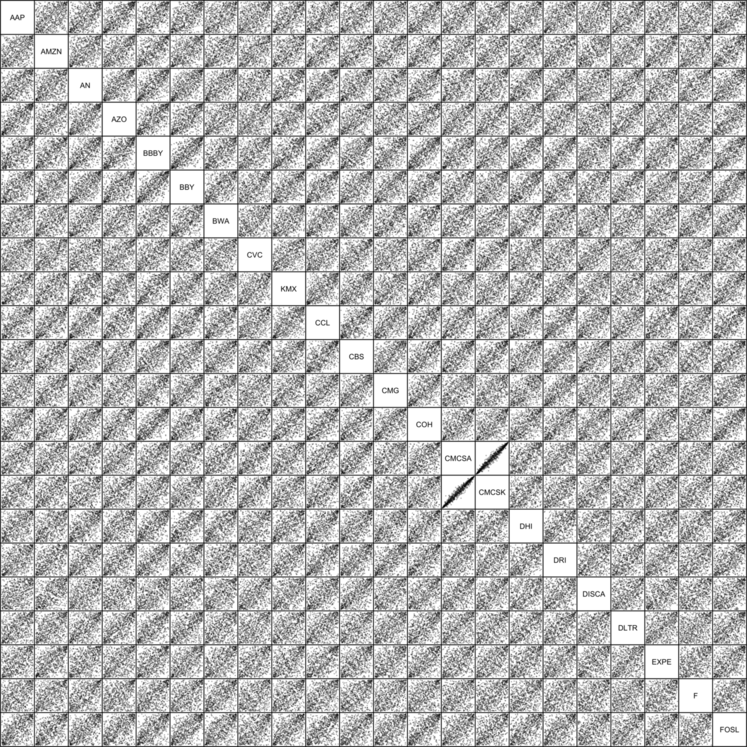}%
  \caption{Scatterplot matrix of $(\bm{U}_{t,j})_t$,
    $j\in\{1,\dots,22\}$. Small displays highlight low spatial frequency
    structure, and hence dependence structure, and allow the dependence of many
    pairs to be assessed and compared simultaneously.}
  \label{fig:SP500:splom:pobs}
\end{figure}

Figure~\ref{fig:SP500:splom:pobs}
displays the scatterplot matrix of the pseudo-observations of the standardized residuals from
marginally fitting the first $22$ constituents of our \SP\ data.  In addition to
being able to simultaneously display and compare many plots at once, the
scatterplot matrix has another important characteristic when it comes to
assessing dependence -- it produces small scatterplots.  Small scatterplots make
our visual system focus on the low spatial frequency characteristics of each
plot and this is ideal for detecting dependence structure from uniforms.

For example, consider again the plots of
Figure~\ref{fig:SP500:plots:pobs}. The right-most three plots appear in
Figure~\ref{fig:SP500:splom:pobs} as the first three plots from the top left
along the first diagonal above the main diagonal; that is, considering the
scatterplot matrix as a $22 \times 22$ matrix, the right-most three plots of
Figure~\ref{fig:SP500:plots:pobs} appear in cells $(1,2)$,
$(2,3)$, and $(3,4)$, respectively, of the scatterplot matrix in Figure~\ref{fig:SP500:splom:pobs}. Moving your eye
along the first three plots of this diagonal, it is easier to see the dependence
in each plot and that this dependence is increasing as the eye travels down and
to the right.  The same visual effect can be achieved with
Figure~\ref{fig:SP500:plots:pobs} by squinting when observing
the plots, or by physically moving farther away (and hence making them smaller).

Note that some care has been taken in the construction of the scatterplot
matrix.  No superfluous annotation appears (no axes, etc.), each point is
plotted with a very small size, and the colour used for plotting
includes an alpha level chosen so that overplotted points will show darker
(so-called \emph{alpha-blending}) to better visually suggest the density of the
distribution.

Looking over the scatterplot matrix one can assess the (in)dependence of any
pair and compare the strength of the dependence for different pairs.  For
example, considering only the top left $4 \times 4$ block, we can see that
position $(1, 4)$ shows the highest dependence in this block, with a
preponderance of points in having high joint returns. This scatterplot is that
of \texttt{AZO} and \texttt{AAP} and shows a stronger dependence than that of
\texttt{AZO} and \texttt{AN} shown in $(3, 4)$ or the right-most plot in
Figure~\ref{fig:SP500:plots:pobs}. In the rest of the matrix there are other
stronger, and many weaker, types of dependence that can also be seen. The
dependence that stands out most is that of \texttt{CMCSA} and \texttt{CMCSK},
near the diagonal in position $(14,15)$. These are the pseudo-observations for
the standardized residuals on two different classes of Comcast shares, which
explains such a strong dependency.

\subsection{Zenplots}\label{sec:zenplots}
A major drawback of the scatterplot matrix is its wasted space.  Every plot
appears twice, once above the main diagonal and once below -- for example, the two
strongest dependencies seen near the main diagonal of
Figure~\ref{fig:SP500:splom:pobs} are the same pairs of Comcast constituents.
In this display of 462 scatterplots, only 231 show different pairs (possibly after
a rotation or reflection).

\begin{figure}[htbp]
  \centering
  \includegraphics[width=\textwidth]{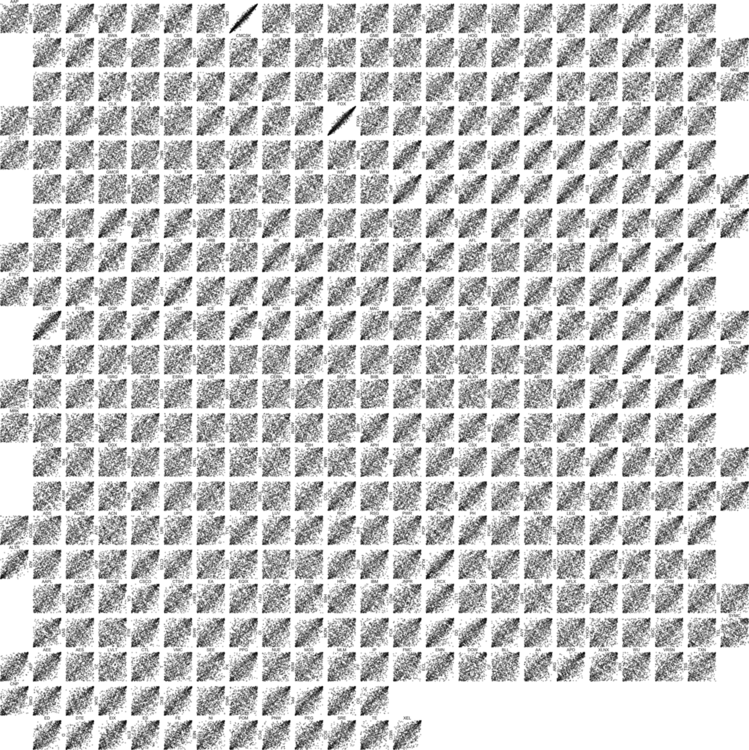}%
  \caption{A zenplot of the pseudo-observations showing the pairs $(1,2), (2,3),\dots, (d-1,d)$ for $d=465$.}
  \label{fig:SP500:zenplot:pobs}
\end{figure}

In contrast, a \emph{zenplot} (or \emph{zigzag expanded navigation plot}) lays
out many more scatterplots by making better use of the space.  The plots are
laid out following a well defined path that zigzags across the page.
Figure~\ref{fig:SP500:zenplot:pobs}
shows a zenplot that displays $d-1=464$ \emph{different} scatterplots in
approximately the same space; to this end we used the function \texttt{zenplot()} of the \R\ package
\texttt{zenplots}. As with the scatterplot matrix one sees two very strong
dependencies.  Unlike the scatterplot matrix, these are two different pairs of
stocks: the one in the first row is that of the Comcast shares as before, but
the one in the fourth row is a pair of two different classes of Twenty-first
Century Fox shares (\texttt{FOX} and \texttt{FOXA}).  The zenplot allows visual
search and comparison over twice as many plots as does a scatterplot matrix in
the same space; note that the labels of the pairs can be made more visible by
zooming in on Figure~\ref{fig:SP500:zenplot:pobs}.

The zenplot lays out the scatterplots as follows.  The first is placed in the
top left corner.  This has variate 1 as its horizontal axis and variate 2 as its
vertical.  The next plot is placed at the right of the first with variate 2 as
its vertical axis and variate 3 as its horizontal.  The third scatterplot is
placed below the second with variate 3 as its horizontal and variate 4 as its
vertical.  The next scatterplot is placed to the right of this sharing the same
vertical axis and having variate 5 as its horizontal axis.  Then the next plot
is placed above the fourth with horizontal variate 5 and vertical variate 6.
One zigzag pattern is now completed and we are in a position similar to the
starting position.  This continues left to right across the page until the end
where the zigzag moves down and reverses its direction to move from right to left.  And
so on until all plots have been laid out.  Whenever an axis is shared, the name
of that variate appears between the two plots.

If one imagines the $d \times d$ scatterplot matrix of all $d$ variates, the
zenplot of Figure~\ref{fig:SP500:zenplot:pobs} has plotted all scatterplots that
lie along the diagonal of the scatterplot matrix immediately above the main
diagonal -- it plots the pairs $(1,2), (2,3),\dots, (d-1,d)$.  For example, the
three scatterplots in the top left corner of Figure~\ref{fig:SP500:zenplot:pobs}
are exactly the three right-most scatterplots of
Figure~\ref{fig:SP500:plots:pobs}. Many other customizations of zenplots are
possible; see \texttt{?zenplot}.

\subsection{Zenpaths}\label{sec:zenpath}
Although a zenplot shows twice as many distinct pairs as does a scatterplot
matrix in the same space, there are still a great many pairs that could be
shown.  For $d=465$ dimensions there are $\binom{d}{2}=$107,880 distinct
plots.
Of these, the zenplot of Figure~\ref{fig:SP500:zenplot:pobs} showed a
little fewer than 500 or about 0.5\% of all that are possible. To look at all plots would require
more than 200 zenplots of the size of that shown in
Figure~\ref{fig:SP500:zenplot:pobs} (or more than 400 scatterplot matrices); see
also our comment in the last paragraph of Section~\ref{sec:model:check} where we
mention the exercise of laying out and actually examining all distinct pairs in a single zenplot.

Better would be to show only the \emph{most interesting} pairs of plots. Here
again the zenplot layout has an enormous advantage over the scatterplot matrix,
simply by not being restricted to a matrix layout.  Instead, the zenplot
displays a particular path (or series of paths) through the scatterplot matrix.

To see this, imagine the $d \times d$ scatterplot matrix.  A path can be
followed through this matrix beginning at any plot and jumping to any other plot
in the same row or column, thus sharing either a vertical or horizontal axis,
and continuing in this fashion to produce a path of any desired length; see
\cite{oldford2011graphs}.  A \emph{zenpath} is any such path which alternates
between searching along a row and along a column (that is, it is a 3d transition path in the sense
of \cite{oldford2011graphs}).

A (default) zenplot display lays out the scatterplots in the order in which the
variates appear in the dataset.  Figure~\ref{fig:SP500:zenplot:pobs}, for
example, follows the variate order $1, \ldots, d$ which corresponds to the
zenpath starting at the top left corner of the scatterplot matrix of all $d$
variates and zigzagging alternating right (same row) and downward (same column)
crossing the variate name on the diagonal to the next plot each time.  The path
ends when the bottom right corner is reached.
To follow any particular zenpath, then, the variate order in the dataset is simply changed to that of the desired zenpath before zenpath is called.

When a measure of interestingness can be assigned to each scatterplot, we
might restrict our displays to show only those plots of high interest.  These can be
laid out as a zenpath (or series of disconnected zenpaths) following the graph
theoretic methods of \cite{hurley2010pairwise, oldford2011graphs} and automated
using the \R\ package \texttt{PairViz} of \cite{hurley2011PairViz}.

For financial data like the \SP, interest often lies in the dependence between
the negative log-returns of any two stocks. Two common dependence measures for
such data are Kendall's tau and Spearman's rho (both are measures of concordance
in the sense of \cite{scarsini1984}) on the standardized residuals. Or, as in certain
quantitative risk management applications, we might be interested primarily in
what happens in the extremes, that is in some measure of tail dependence between
the two returns. In what follows, we focus on the latter case.

\subsubsection{Tail dependence based on pairwise fitted $t$ copulas}
\label{sec:taildep:biv:t}
Here our measure of  ``interestingness'' will be a formal measure of the (upper) tail dependence between each pair of
negative log-returns by fitting bivariate copulas.
To be concrete, for any $(Z_1, Z_2)$ with joint distribution function $H$,  continuous marginal distribution
functions $F_1,F_2$, and corresponding marginal quantile functions $Q_1, Q_2$, a measure of upper tail dependence, for any $p \in [0,1]$, can be
taken to be $\P(Z_2>Q_2(p)\,|\,Z_1>Q_1(p))$.  This probability is symmetric in $Z_1$ and $Z_2$ and  the larger it is,
the greater is the dependence between $Z_1$ and $Z_2$ in the top right corner
of the bivariate distribution. Taking
its limit as $p \rightarrow 1$, we have the
\emph{coefficient of upper tail dependence}
\begin{align*}
  \lambda = \lim_{p \uparrow 1} \P(Z_2>Q_2(p)\,|\,Z_1>Q_1(p)).
\end{align*}
This is often of interest when modelling joint high quantile
exceedances in quantitative risk management; see, for example, \cite[Example~7.40]{mcneilfreyembrechts2015}.
If the bivariate copula for $H$ is $C$, then it is easy to show that
\begin{align}
  \lambda = \lim_{p\uparrow 1}\frac{1-2p+C(p,p)}{1-p}\label{lambda}
\end{align}
and so the coefficient depends only on the underlying copula $C$ of $H$ and not on
the margins $F_1,F_2$.

Although non-parametric estimation of $\lambda$ is a possibility it requires
deciding on how best to choose the number of data points $k = np$ or
equivalently the probability level $p$ on which to base estimation.  For our
purposes, it is enough to fit parametric copula models to the
pseudo-observations~\eqref{pobs} of each pair of variables and the implied
tail-dependence coefficients be computed (based on equation \eqref{lambda}).

In higher dimensions practitioners have begun working with matrices
$\Lambda = [\lambda_{i j}]$ of pairwise upper tail-dependence coefficients;
see \cite{embrechtshofertwang2016} for a characterization and practical
application.
Estimates of the entries of $\Lambda$ will be taken to be the pairwise estimates $\widehat{\lambda}_{ij}$.

To this end we will use bivariate $t$ copulas and fit them using the approach of
\cite{mashalzeevi2002} (see also \cite{demartamcneil2005}) and the recently
available implementation \texttt{fitCopula(, method = "itau.mpl")} from the \R\
package \texttt{copula} of \cite{copula}. In terms of the fitted degrees of
freedom $\nu$ and the correlation $\rho$ (the off-diagonal element of the
bivariate $t$ copula's scale matrix $P$), the upper tail-dependence coefficient
$\lambda$ of a $t$ copula is
\begin{align}
  \lambda = 2t_{\nu+1}\biggl(-\sqrt{\frac{(\nu+1)(1-\rho)}{1+\rho}}\biggr),
  \label{eq:lamFromRho}
\end{align}
where $t_{\nu+1}(x)$ is the distribution function of the $t$ distribution on $\nu+1$ degrees of freedom evaluated at $x$.
Replacing the parameters by their estimates gives an estimate for $\lambda$.

\begin{figure}[htbp]
  \centering
  \includegraphics[width=0.45\textwidth]{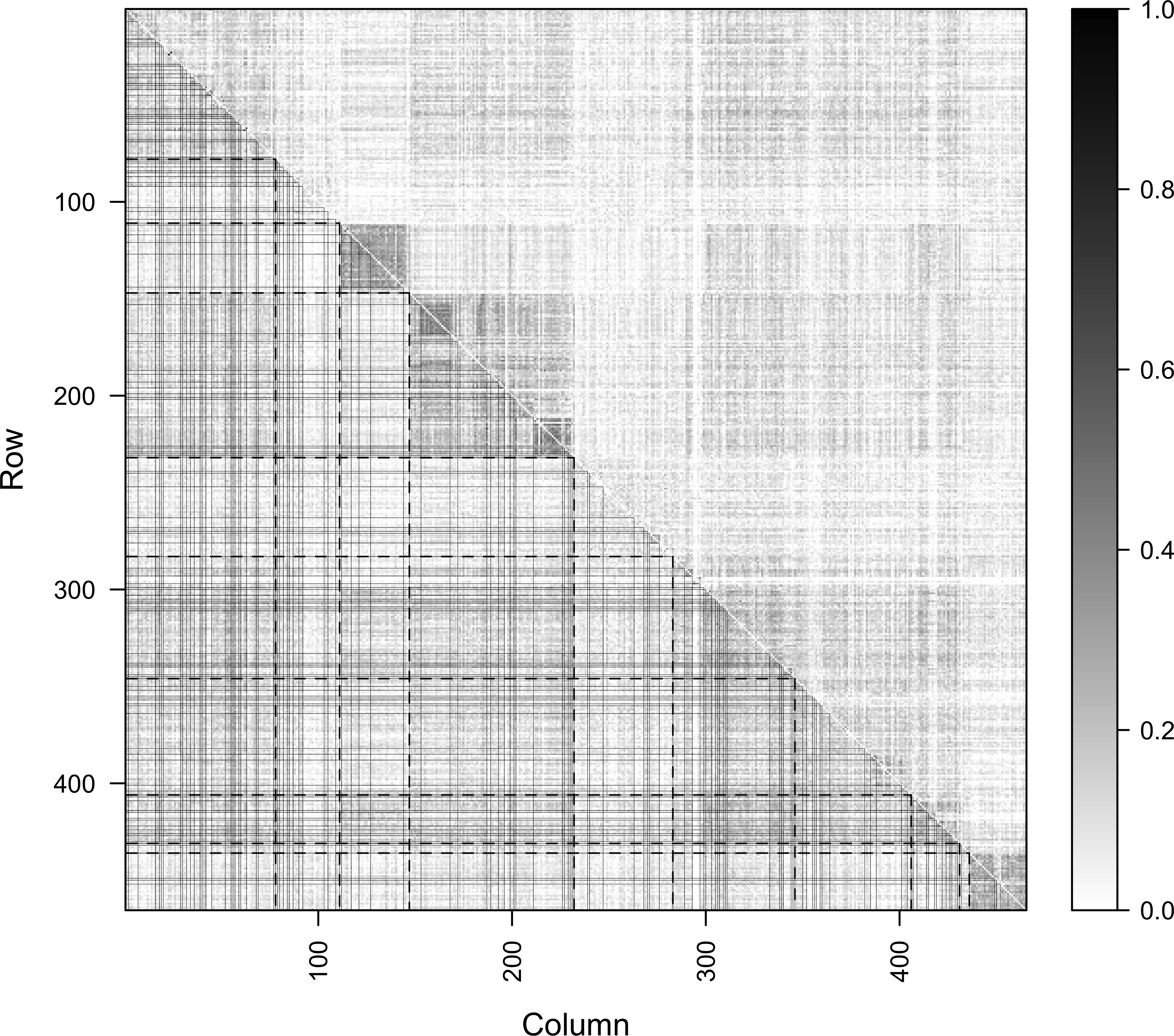}%
  \hfill
  \raisebox{4mm}{\includegraphics[width=0.51\textwidth]{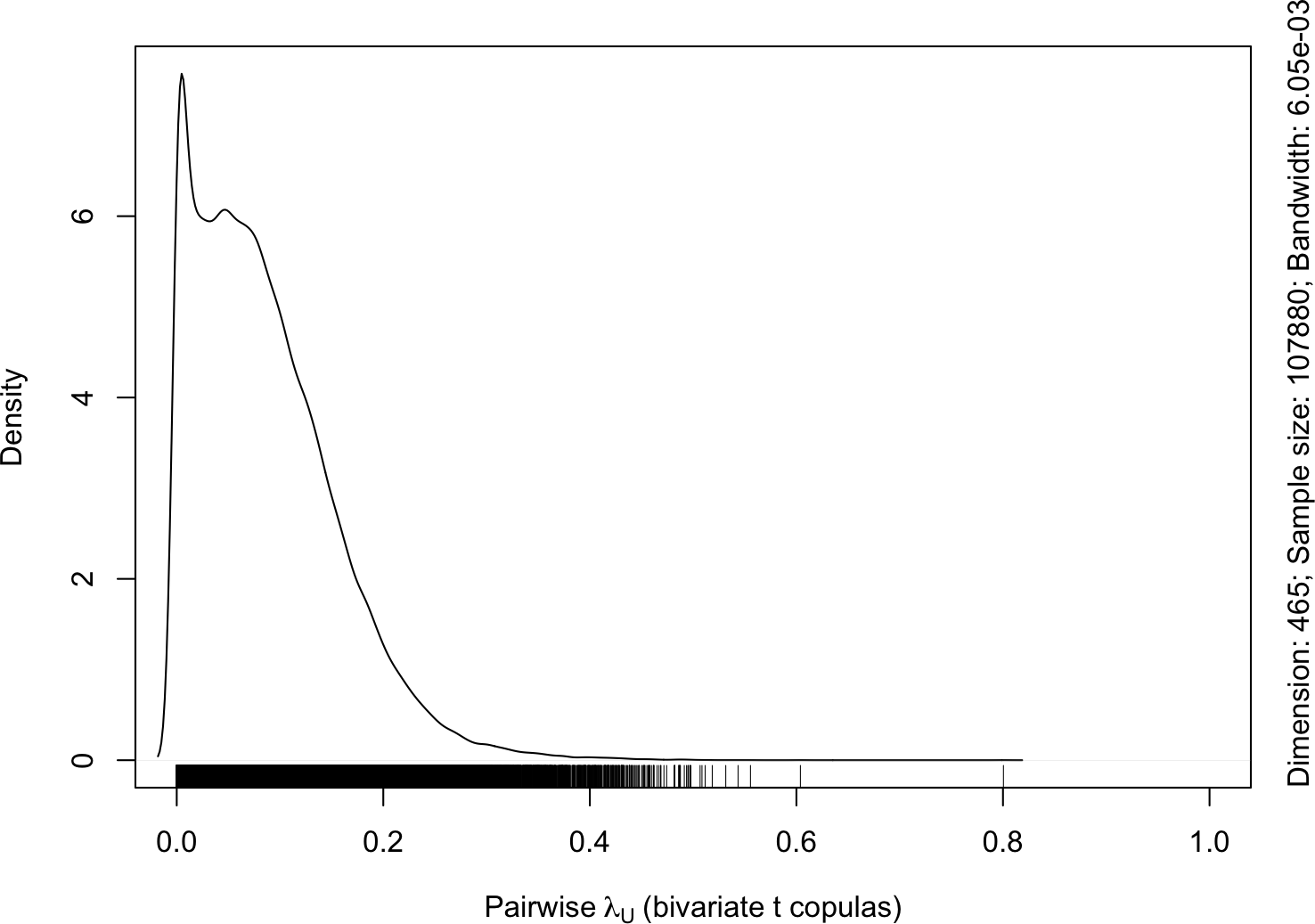}}%
  \caption{Pairwise estimated upper tail-dependence coefficients $\lambda$ from
    fitted bivariate $t$ copulas (different degrees of freedom).  At left is the
    entire matrix $\Lambda$ with GICS sectors (dashed lines) and sub-sectors
    (finer solid lines) indicated; at right is a plot of the estimated density
    of all $\binom{d}{2}$ pairwise entries of $\Lambda$.}
  \label{fig:SP500:Lambda}
\end{figure}

Figure~\ref{fig:SP500:Lambda}
shows pairwise estimated upper tail-dependence coefficients $\lambda$ for all
$\binom{d}{2} = 107,880$ pairs. At left, these are displayed in a matrix
arrangement analogous to a scatterplot matrix except that, instead of a
scatterplot, in each cell the value of the corresponding $\lambda$ estimate is
shown using a grey-scale encoding of $[0,1]$.  This is essentially the pairwise estimated
matrix $\Lambda$.  At right, the overall density
of all pairwise $\lambda$ estimates is shown.  Most values, for example, are
less than 0.3.

The rows (columns) of the estimated matrix $\Lambda$ are arranged in
Figure~\ref{fig:SP500:Lambda} so that stocks in the same GICS sector, and within
each sector in the same sub-sector, appear next to each other in the same row
(column).  The boundaries between sectors are marked by dashed lines appearing
below the diagonal.  The sectors appear top to bottom (left to right) in
alphabetical order as before, namely: ``Consumer discretionary'' (78
constituents = first 78 rows), ``Consumer staples'' (33), ``Energy'' (36),
``Financials'' (85), ``Health care'' (51), ``Industrials'' (63), ``Information
technology'' (60), ``Materials'' (25), ``Telecommunications services'' (5), and
``Utilities'' (29).  Fine lines below the diagonal mark the sub-sectors within
each sector (more easily discerned by zooming in on the display).

Even with such low estimated values for the upper tail-dependence coefficients,
some structure is revealed by the estimated matrix $\Lambda$ of
Figure~\ref{fig:SP500:Lambda}.  The darkest blocks along the diagonal correspond
to stocks in the same sector and within the same sub-sector.  The third diagonal
block down, for example, shows relatively stronger tail dependencies existing
between constituents of the ``Energy'' sector.  Such information can, for
example, be used to set up a hierarchical dependence model with
within-sector dependencies modelled differently than might be any between-sector dependencies.  Off the
diagonal, the very light region immediately above the dark diagonal ``Energy''
block suggests overall little dependence between the constituents of the ``Energy''
sector and those of the ``Consumer staples'' sector.  The cell values of
$\Lambda$ can thus help determine which corresponding scatterplots of
pseudo-observations to examine more closely along some zenpath(s) of interest.

\begin{figure}[htbp]
  \centering
  \includegraphics[width=0.7\textwidth]{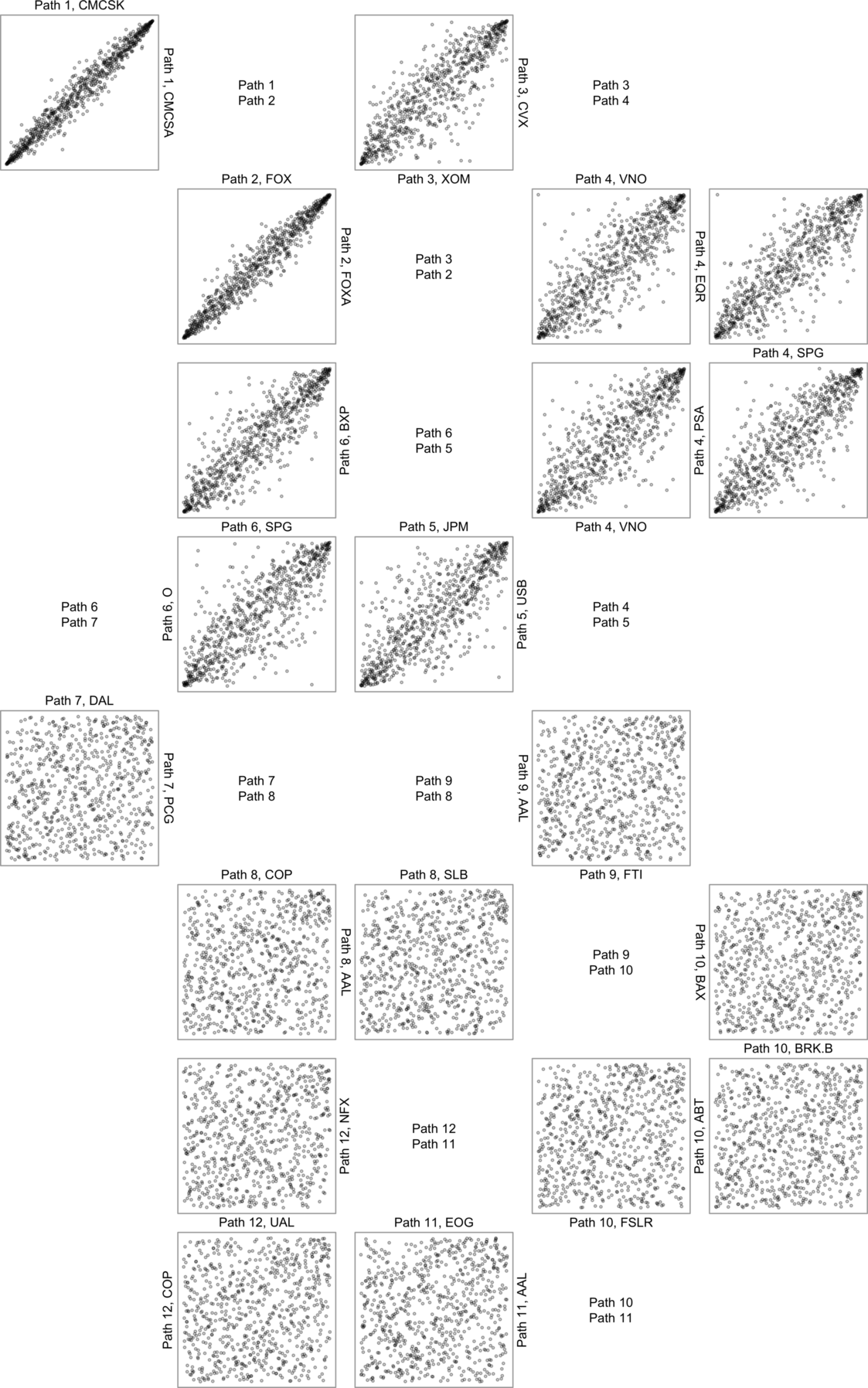}%
  \caption{A zenplot constructed from a zenpath displaying the pseudo-observations of
  those 10 pairs of variables with largest (first) and then those 10 pairs with
  smallest (last) upper tail-dependence coefficient. When variables along the path are
  connected, they are displayed in one contiguous zenpath, abbreviated by
  Path~1--Path~12. Concatenated paths are separated by a block marking the
  transition from one path to another.}
  \label{fig:SP500:zenplot:extreme:lambdas}
\end{figure}

Finding interesting zenpaths from $\Lambda$ has been automated by the function
\texttt{zenpath()}.
In Figure~\ref{fig:SP500:zenplot:extreme:lambdas}
\texttt{zenpath()} was used to find the ten pairs having strongest upper tail
dependence and the ten pairs having weakest upper tail dependence.  All twenty
plots are arranged in order from highest upper tail dependence to lowest upper
tail dependence.  The zigzag pattern is followed exactly as in
Figure~\ref{fig:SP500:zenplot:pobs} except now there are occasional places
showing only a couple of labels where a scatterplot might have been expected.
The reason for this is that these places mark boundaries where one zenpath ends
and another one begins.  Recall that a zenpath connects plots which share a
variate (by alternating row and column selections in a scatterplot matrix).
When no such variate is shared the zenpath ends and a new one begins (the change
being equivalent to moving from one cell in a scatterplot matrix to another in a
different row \emph{and} different column). The zenplot of
Figure~\ref{fig:SP500:zenplot:extreme:lambdas} is the concatenation of twelve
different zenpaths.

The two strongest upper tail-dependencies are those we have already seen, namely
the two classes of Comcast shares and the two classes of Twenty-first Century
Fox shares.  Because these two pairs share no variates, they define the first
two zenpaths (each of length one plot, or two variates) and are separated by the
space labelling the end of path 1 and the beginning of path 2.  The third path,
also of length one, consists of the pair of stocks from the ``Energy'' sector,
\texttt{CVX} (Chevron Corp.) and \texttt{XOM} (Exxon Mobil Corp.).  Next is a
path of length four, consisting of the variates \texttt{VNO} (Vornado Realty
Trust), \texttt{EQR} (Equity Residential), \texttt{SPG} (Simon Property Group
Inc), and \texttt{PSA} (Public Storage), all from the ``Financials'' sector.
The stock \texttt{VNO} appears in two of these four pairs.  Path 5 consists of
the two ``Financials'' sector stocks \texttt{USB} (U.S. Bancorp) and
\texttt{JPM} (JPMorgan Chase \& Co.) and path 6 of three more, \texttt{BXP}
(Boston Properties), \texttt{SPG} (Simon Property Group Inc), and \texttt{O}
(Realty Income Corporation).  Note that every pair of variates, of the ten
having the strongest upper tail-dependence coefficient, is formed from stocks
from the same GICS sector and no variate turns up paired with another from a different sector.

The variate pairs with weakest upper tail dependence appear in the bottom half
of Figure~\ref{fig:SP500:zenplot:extreme:lambdas} and are essentially
indistinguishable from uniform scatterplots (see
Figure~\ref{fig:SP500:plots:pobs}(a)).  This suggests that these pairs of standardized
residuals can be considered to be independent. There are again six
paths: path 7 \texttt{DAL} (Delta Air Lines) from the ``Industrials'' sector,
\texttt{PCG} (PG\&E Corp.) from the ``Utilities'', path 8 \texttt{COP}
(ConocoPhillips) ``Energy'', \texttt{AAL} (American Airlines Group)
``Industrials'', and \texttt{SLB} (Schlumberger Ltd.) ``Energy'', path 9
\texttt{AAL} (American Airlines Group) ``Industrials'', and \texttt{FTI} (FMC
Technologies Inc.)  ``Energy'', path 10 \texttt{BAX} (Baxter International Inc.)
``Health Care'', \texttt{BRK.B} (Berkshire Hathaway) ``Financials'',
\texttt{ABT} (Abbott Laboratories) ``Health Care'', and \texttt{FSLR} (First
Solar Inc) ``Information Technology'', path 11 \texttt{AAL} (American Airlines
Group) ``Industrials'', and \texttt{EOG} (EOG Resources) ``Energy'', and finally
path 12 \texttt{NFX} (Newfield Exploration Co) ``Energy'', \texttt{UAL} (United
Continental Holdings) ``Industrials'', and \texttt{COP} (ConocoPhillips)
``Energy''.  Note that the ten pairs of variates with weakest upper tail dependence
are from different GICS sectors.

\begin{figure}[htbp]
  \centering
  \includegraphics[width=0.7\textwidth]{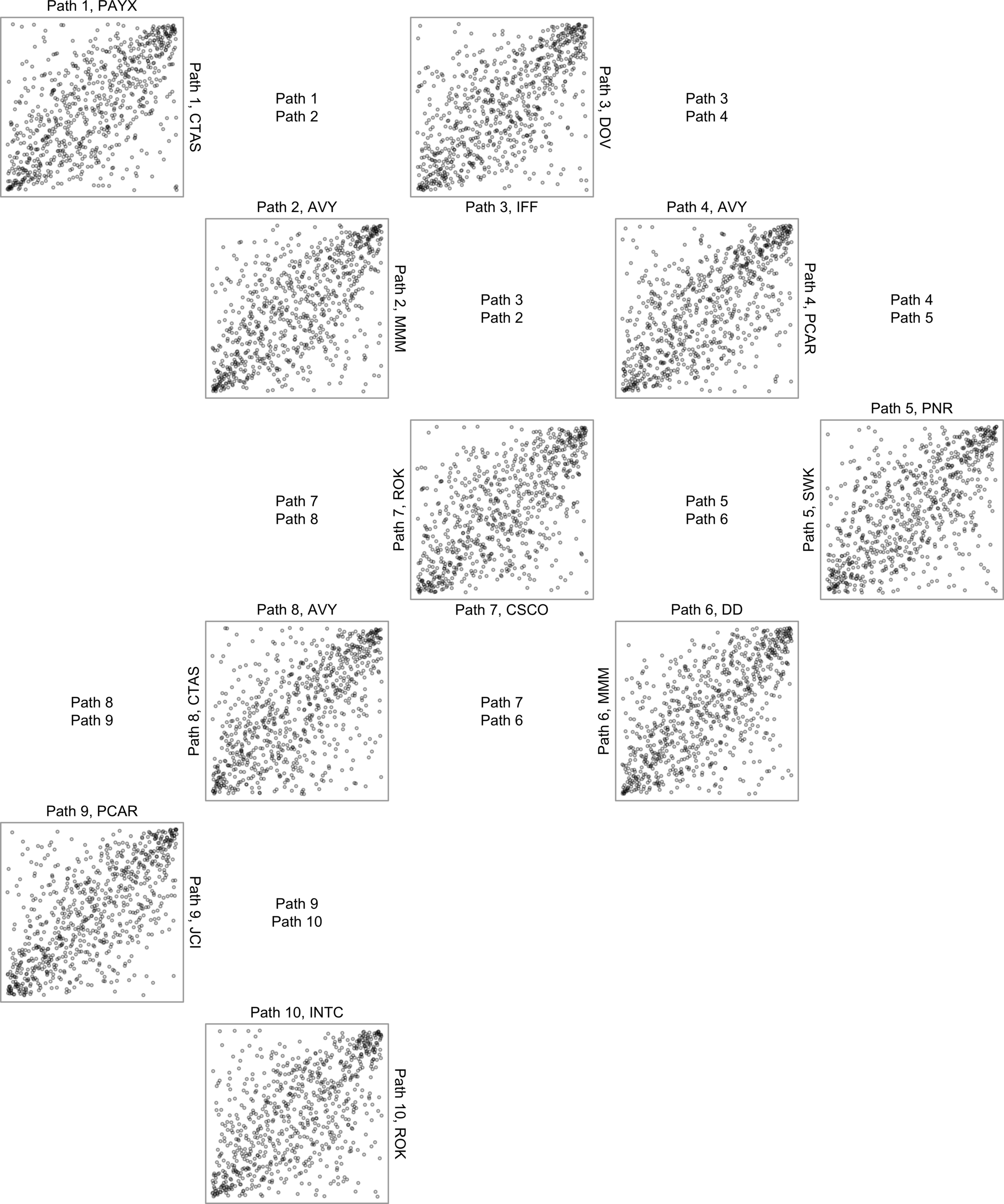}%
   \caption{A zenplot constructed from a zenpath displaying the
    pseudo-observations of those 10 pairs of variables with largest upper tail-dependence
    coefficient which belong to different GICS business sectors.}
  \label{fig:SP500:zenplot:extreme:lambdas:cross}
\end{figure}

Given that the weakest dependencies occur between constituents from different
GICS sectors, one might ask what pairs of constituents from different sectors
have the strongest upper tail dependencies. To answer this,
Figure~\ref{fig:SP500:zenplot:extreme:lambdas:cross}
shows the ten such cross sector pairs having greatest tail dependence in
descending dependence order.  An examination of the individual constituents
reveals that every pair has one stock from the ``Industrials'' sector most
frequently paired with either a ``Materials'' one (5 pairs), or with an
``Information Technology'' stock (3 pairs).  Furthermore, in all ten pairings an
examination of the companies suggests that the dependence is not really
surprising -- for example, the strongest is between \texttt{PAYX} a provider of human
resource outsource services from the ``Information technology'' sector and
\texttt{CTAS} from the ``Industrials'' sector that provides such ``diversified
support services'' as corporate identity, promotional services, restroom
cleaning, and supplies, and the weakest dependence of the ten is between Intel
Corp. (\texttt{INTC}, ``Information Technology'') and Rockwell Automation
Inc. (\texttt{ROK}, ``Industrials'').

Note that there are as many paths in
Figure~\ref{fig:SP500:zenplot:extreme:lambdas:cross} as there are pairs
displayed.  Although some constituents appear more than once (for example,
\texttt{AVY} Avery Dennison Corp. (3), \texttt{CTAS} Cintas Corporation (2),
\texttt{MMM} 3M Company (2), \texttt{PCAR} PACCAR Inc. (2), and \texttt{ROK}
Rockwell Automation Inc. (2)), nowhere along the zenpath of decreasing upper
tail dependence is one of the repeated constituents shared by consecutive plots.

Looking again at Figure~\ref{fig:SP500:zenplot:extreme:lambdas}, we see
that only three sectors appear in the top ten strongest dependencies, namely
``Consumer discretionary'', ``Energy'', and ``Financials''.  This misses
seven out of the ten GICS sectors.  We might ask, then, what pair of
constituents within each and every sector has the largest upper tail dependence?

\begin{figure}[htbp]
  \centering
  \includegraphics[width=0.7\textwidth]{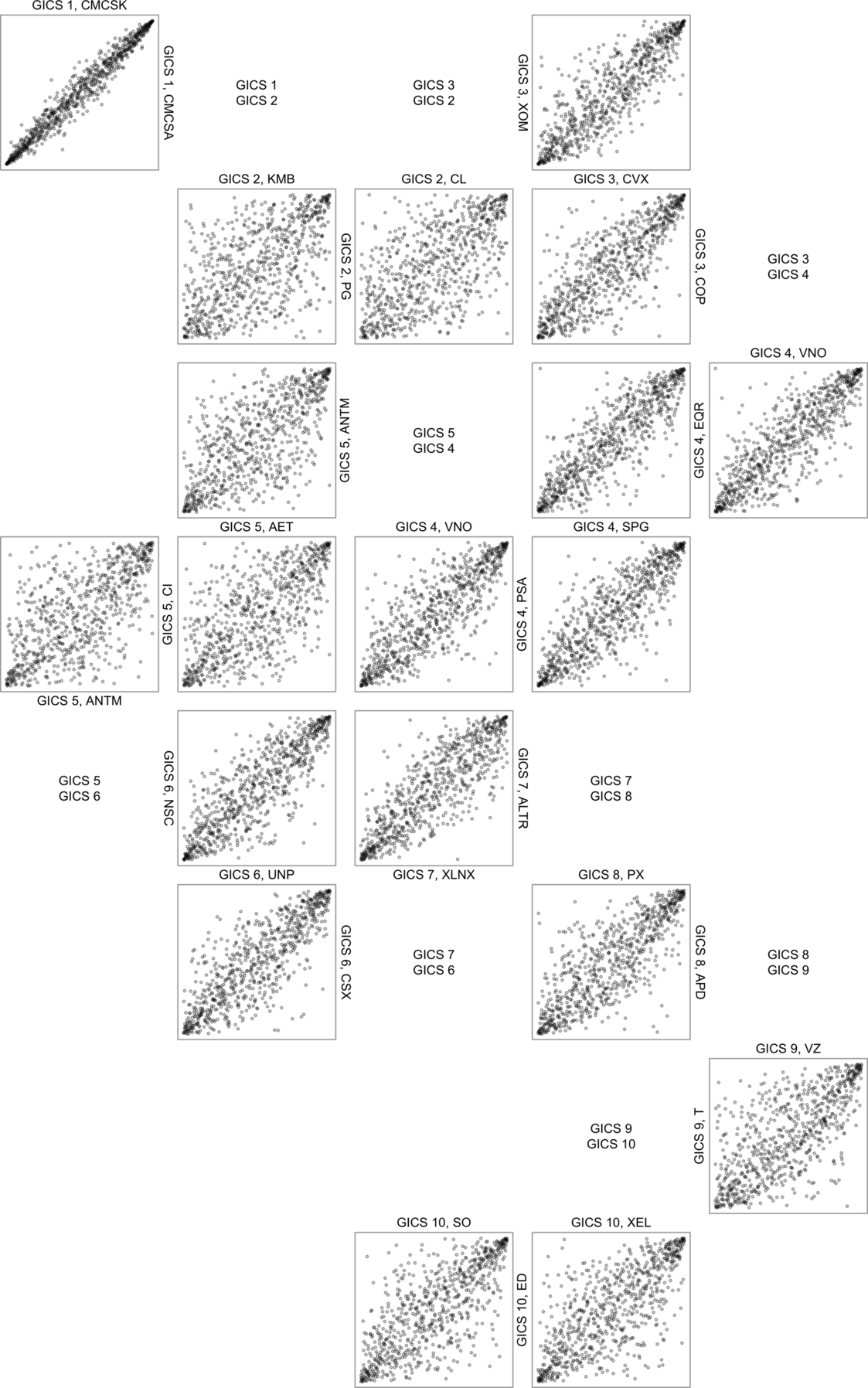}%
  \caption{A zenplot constructed from a zenpath displaying the
    pseudo-observations of those groups of connected pairs of constituents with
    largest upper tail-dependence coefficient within each GICS business sector; the latter (in
    alphabetical order) are labelled GICS~1--GICS~10.}
  \label{fig:SP500:zenplot:extreme:lambdas:sector}
\end{figure}

Figure~\ref{fig:SP500:zenplot:extreme:lambdas:sector}
shows a zenplot of the ten GICS sectors (in alphabetical order as before) as ten
separate zenpaths labelled GICS~1--GICS~10.  Within each sector, the constituent
pairs are sorted from greatest to least dependence.

The first plot in the path of each sector in
Figure~\ref{fig:SP500:zenplot:extreme:lambdas:sector} displays the
pseudo-observations for that pair of constituents having the largest upper
tail-dependence coefficient within that sector. For each sector, the zenpath
beginning at that pair continues in decreasing tail dependence as long as the
path is connected; it ends as soon as the path ends in that sector.  For
example, within the first GICS sector (``Consumer discretionary'') only one pair
appears, as no connection to the next pair within that sector can be made
according to the zenpath of decreasing tail dependence; this is why the strong
relation between the \texttt{FOX} and \texttt{FOXA} shares of
Figure~\ref{fig:SP500:zenplot:extreme:lambdas} does not also appear in
Figure~\ref{fig:SP500:zenplot:extreme:lambdas:sector}.  However, the second
sector (``Consumer staples'') the two pairs with strongest tail dependence share
the variate \texttt{PG} (Procter \& Gamble) and so appear as a connected pair
joining three well known companies (from two different sub-sectors) offering
personal and household products (viz.\ \texttt{KMB} (Kimberly-Clark), \texttt{PG},
and \texttt{CL} (Colgate-Palmolive)).

The concatenated zenpaths of
Figure~\ref{fig:SP500:zenplot:extreme:lambdas:sector} separate the groups
visually and give some sense of their size.  The groups, their stocks in order,
and the sub-sectors to which they belong are as follows: GICS~1 or ``Consumer
discretionary'' (\texttt{CMCSK} and \texttt{CMCSA}, both Comcast); GICS~2 or
``Consumer staples'' (\texttt{KMB} (Kimberly-Clark), \texttt{PG} (Proctor \&
Gamble), and \texttt{CL} (Colgate-Palmolive) from two closely related but different
personal or household product sub-sectors); GICS~3 or ``Energy'' (\texttt{XOM}
(Exxon), \texttt{CVX} (Chevron), \texttt{COP} (ConocoPhillip), from two closely
related but different oil \& gas sub-sectors); GICS~4 or ``Financials''
(\texttt{VNO} (Voronado), \texttt{EQR} (Equity residential), \texttt{SPG} (Simon
Property group), \texttt{PSA} (Public Storage), and \texttt{VNO} again, all
``REITs''); GICS~5 or ``Health care'' (\texttt{ANTM} (Anthem Inc.), \texttt{AET}
(Aetna), \texttt{CI} (CIGNA Corp.) , and \texttt{ANTM} again, all ``Managed health
care''); GICS~6 or ``Industrials'' (\texttt{NSC} (Norfolk Southern), \texttt{UNP}
(Union Pacific), \texttt{CSX} (CSX Corp.), all ``Railroads''); GICS~7 or
``Information technology'' (\texttt{XLNX} (Xilinx Inc.), \texttt{ALTR} (Altera
Corp.), both ``Semiconductors''); GICS~8 or ``Materials'' (\texttt{PX} (Praxair),
\texttt{APD} (Air products and chemicals), both ``Industrial gases''); GICS~9 or
``Telecommunication services'' (\texttt{VZ} (Verizon), \texttt{T} (AT\&T), both
``Integrated telecom services''); and GICS~10 ``Utilities'' (\texttt{XEL} (Xcel
Energy Inc.), \texttt{ED} (Consolidated Edison), \texttt{SO} (Southern Co.), two
``Electric utilities'' one ``MultiUtilities'').  The zenplot has clearly picked
out some interesting related groupings with many of the strongest relations
within a sector also appearing within the same (or closely related by
definition) sub-sector.

Clearly some sectors show much weaker within-sector dependence than do others.
We could continue in this way, using zenplots to more deeply explore dependence
within or between sectors and sub-sectors.  Instead, we now turn our attention
to how zenplots might also be used in assessing the models we are using.

\section{Model assessment}\label{sec:model:check}
Zenplots can be used to quickly arrange nearly any visual display in a compact
but informative way. Here we illustrate how zenplots, as well as some other
displays, might be used in the very important task of model assessment.

\subsection{Checking serial dependence of the marginals}\label{sec:model:check:margins:serial}
We modelled each component series of negative log-returns by an
$\ARMA(1,1)-\GARCH(1,1)$ model with serially (or temporally) independent $t$
innovations. The assumption of serial independence might be assessed visually
via the autocorrelation function (ACF) of the standardized residuals
$\widehat{Z}_{t,j}$, $t\in\{1,\dots,T\}$, $j\in\{1,\dots,d\}$.

\begin{figure}[htbp]
  \centering
  \includegraphics[width=\textwidth]{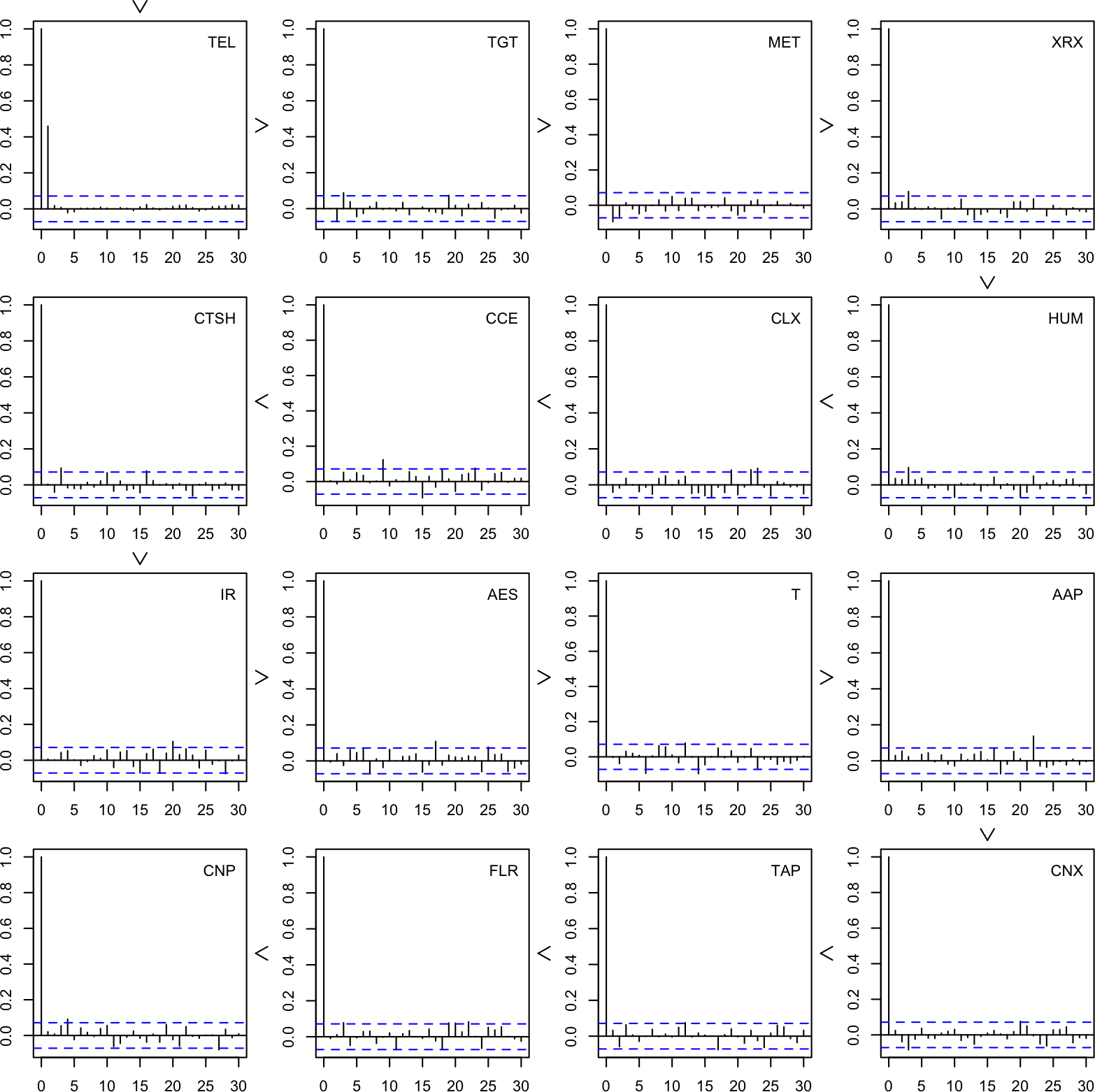}%
  \caption{A zenplot of ACFs of those 16 standardized residual series with
    largest maximal (over lags 1 to 30) Ljung--Box test statistics.}
  \label{fig:SP500:zenplot:residual:check:LB}
\end{figure}

Figure~\ref{fig:SP500:zenplot:residual:check:LB}
shows the ACFs up to lag 30 of sixteen standardized residual series. The order
of these series is chosen to place earliest in the sequence those
  components which most challenge the null hypothesis of serial independence.
  We measure the strength of this challenge as follows.  For each component, the
  Ljung--Box test statistic is computed for every lag from 1 to 30.  Because,
  under the null hypothesis, the magnitude of this statistic increases with the lag, the p-value for that lag's test statistic is used to place the different lag test statistics on a common scale (that is, ordering by test statistic magnitude does not make sense here since the different lags will have different magnitude test statistics each with its own distribution when the hypothesis of serial independence holds).  The smallest of these p-values corresponds to that lag which most greatly challenges the hypothesis of serial independence for this component as measured by a Ljung--Box test.  The components are then ordered from smallest to largest by these minimal p-values; Figure~\ref{fig:SP500:zenplot:residual:check:LB} shows the  ACFs for the first  sixteen (most challenging) components in this order.

\begin{figure}[htbp]
  \centering
  \includegraphics[width=\textwidth]{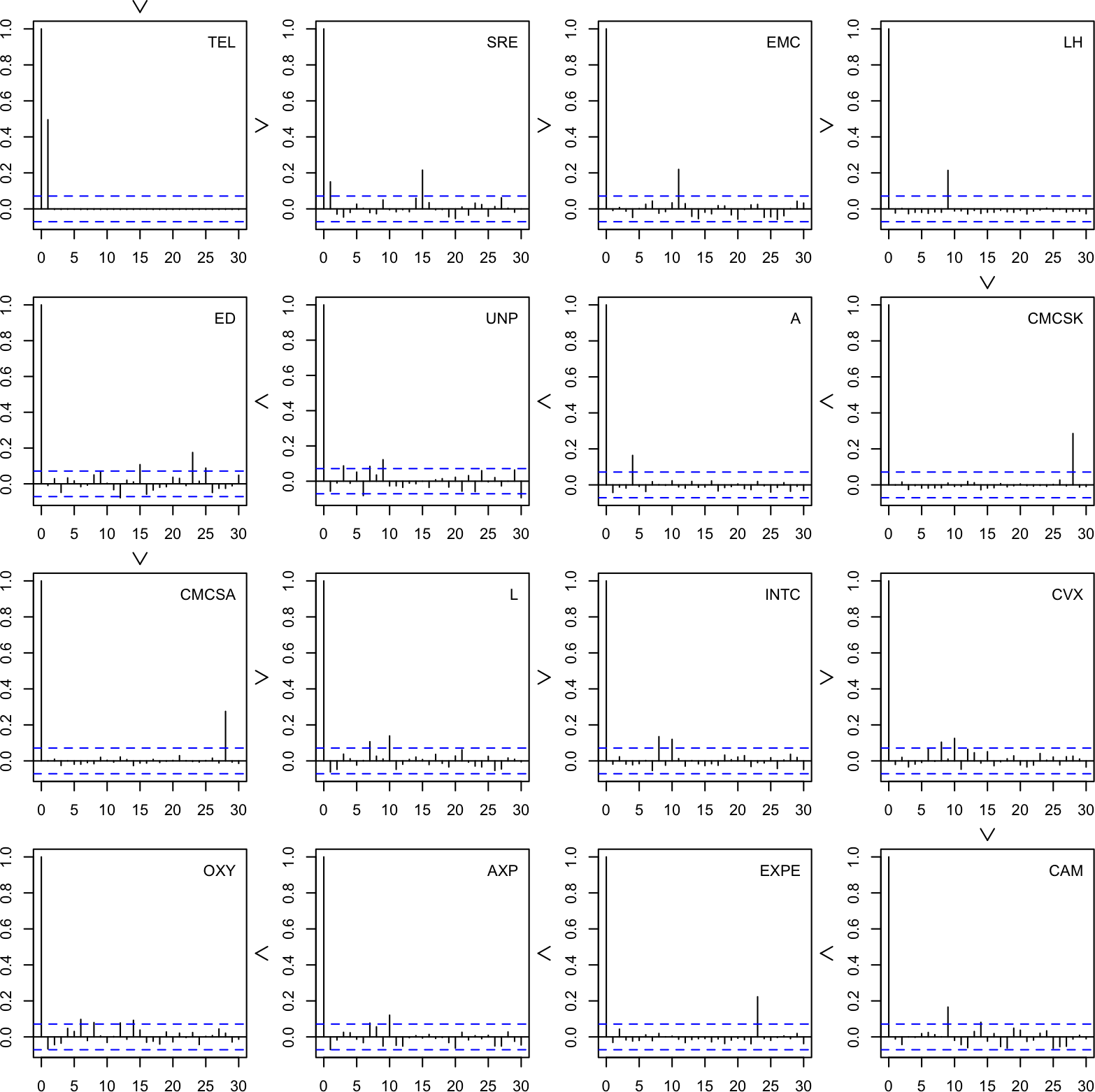}
  \caption{A zenplot of ACFs of those 16 squared standardized residual series
    with largest maximal (over lags 1 to 30) Ljung--Box test statistics.}
  \label{fig:SP500:zenplot:residual:check:LB:squared}
\end{figure}

Another way in which the null hypothesis of serial independence might be
violated is if there were some remaining temporal dependence in the
volatility.  To assess this, we apply the same strategy as above to challenge the null hypothesis of serial independence except that now, since interest lies in volatility, we use the series of squared estimated standardized
residuals $\widehat{Z}_{t,j}^2$, $t \in \{1,\ldots, T\}$, for each component $j$ when conducting the Ljung-Box tests.
Figure~\ref{fig:SP500:zenplot:residual:check:LB:squared}
shows the ACFs for those 16
components $j$ having the smallest p-values over all components and, for each component, over all Ljung--Box tests
for lags 1 to 30. As was the case in Figure~\ref{fig:SP500:zenplot:residual:check:LB}, the component most challenged by possible serial dependence is \texttt{TEL}; all remaining components in Figure~\ref{fig:SP500:zenplot:residual:check:LB:squared}  differ from those which appeared in Figure~\ref{fig:SP500:zenplot:residual:check:LB}.

Together, then, the zenplots of Figures~\ref{fig:SP500:zenplot:residual:check:LB} and
\ref{fig:SP500:zenplot:residual:check:LB:squared} serve up the ACF displays of those components that demonstrate the greatest challenges to serial independence (in level or volatility), at least as measured by the minimum p-value of a Ljung-Box test for every lag from 1 to 30. That is, by this particular measure, in each Figure these are the strongest $16$ cases against the null hypothesis for all $465$ components.
An examination of these ACFs reveals little serial dependence with one notable and jarring exception.  Appearing as the first (and strongest) case in each zenplot, the  ACF of the component \texttt{TEL} (TE Connectivity Ltd.) shows a significant lag 1 auto-correlation where none should exist (since these are the auto-correlations of the estimated standardized residuals from an  $\ARMA(1,1)-\GARCH(1,1)$ model).

\begin{figure}[htbp]
  \centering
  \includegraphics[width=0.48\textwidth]{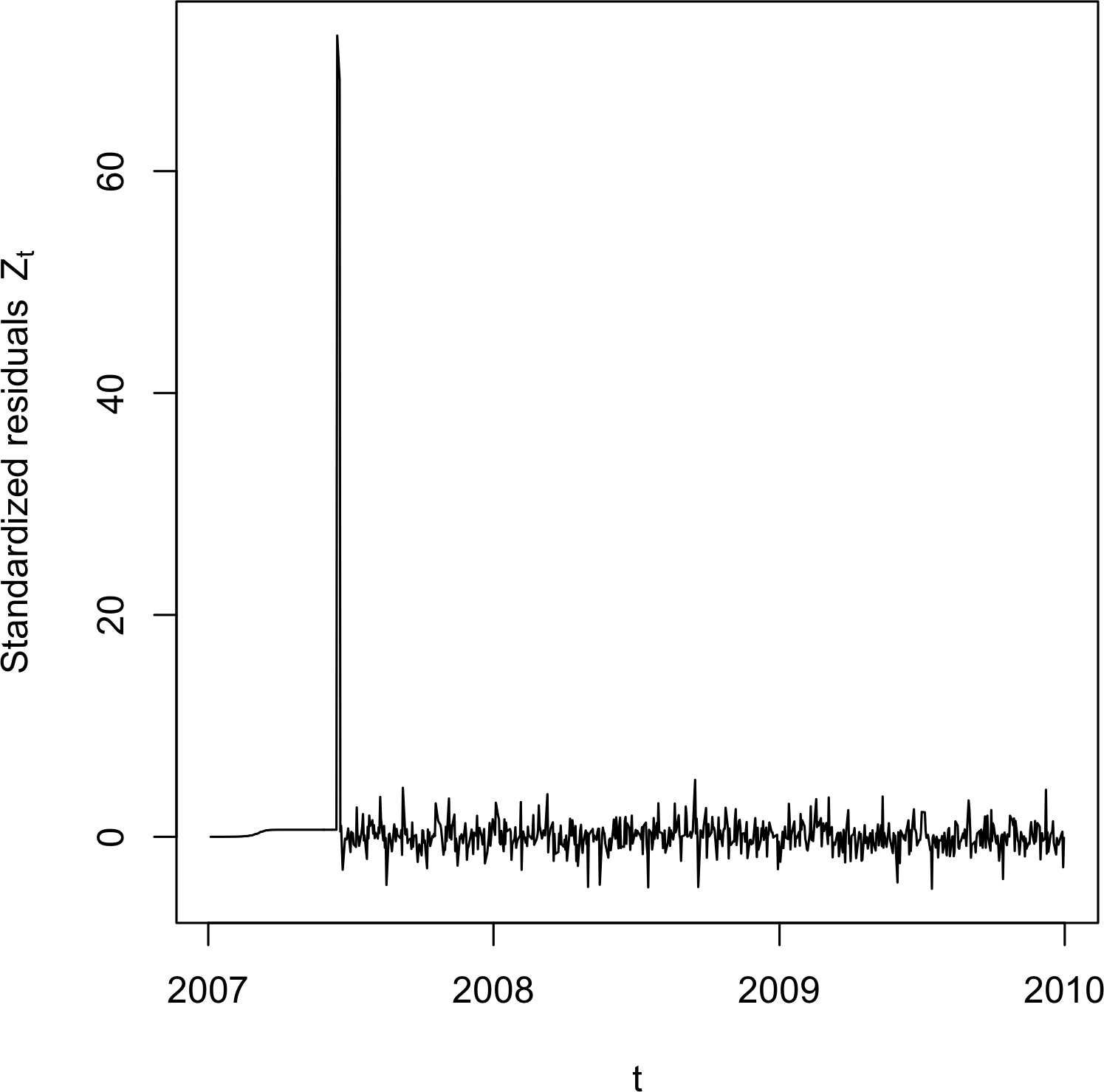}%
  \hfill
  \includegraphics[width=0.48\textwidth]{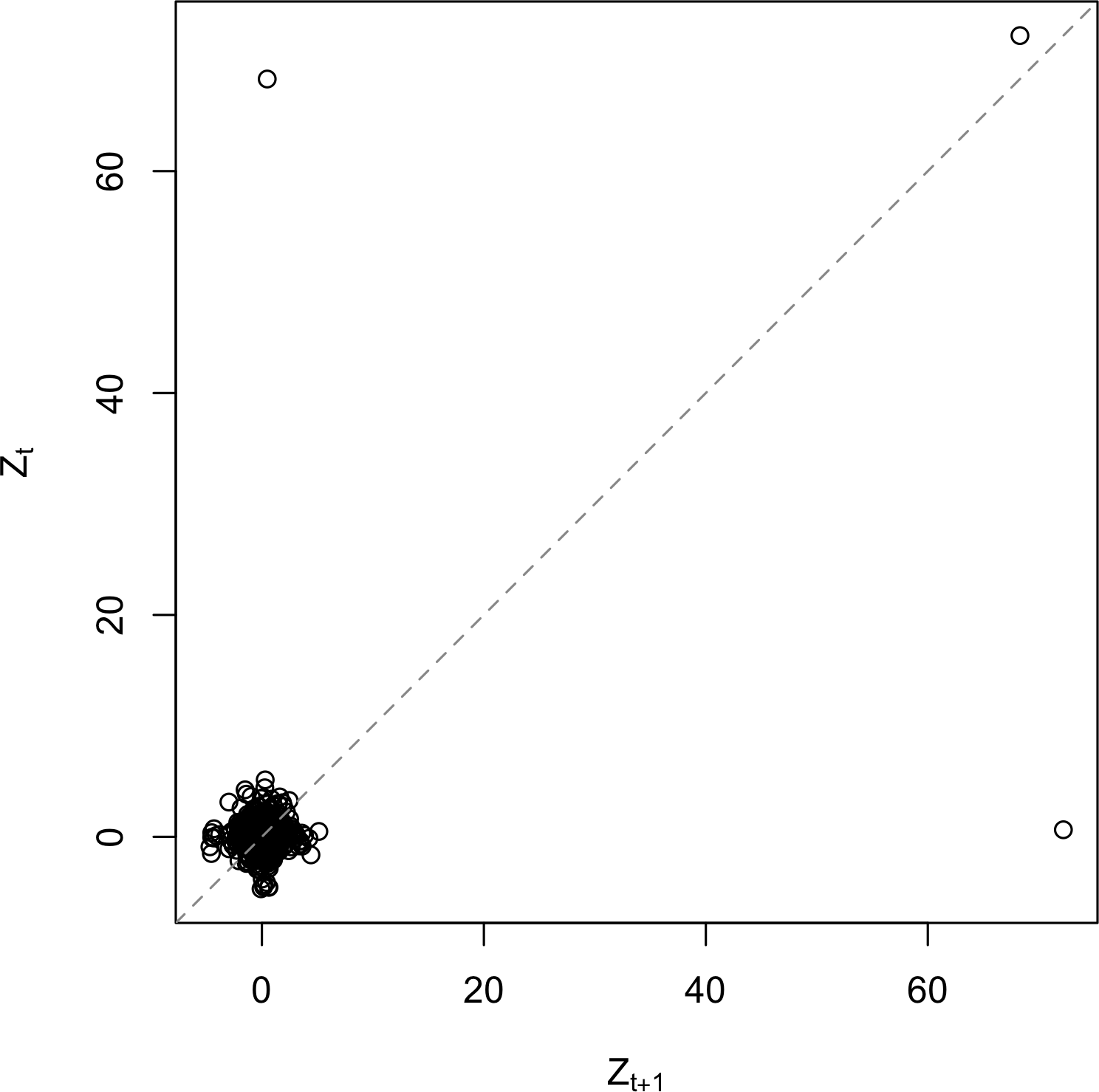}%
  \caption{The standardized residuals from fitting a marginal
$\ARMA(1,1)-\GARCH(1,1)$ model to \texttt{TEL} (TE Connectivity Ltd.) plotted at left over time and at right against the values in the immediately previous time period (a lag 1 plot).}
  \label{fig:SP500:zenplot:residual:check:LB:ACF_problem}
\end{figure}

To see how this might have occurred, in Figure~\ref{fig:SP500:zenplot:residual:check:LB:ACF_problem}
we plot the standardized residuals over time for the component \texttt{TEL} (the left display) together with their lag 1 plot (the right display).  Two adjacent huge positive residuals appear near the beginning of the series.   Their adjacency is perhaps clearer in the lag plot, where a single point appears in the top right corner as a result.  This single point is far enough away from the bulk of (essentially uncorrelated) points in the bottom left of the lag plot to have induced the spurious auto-correlation.  Of course, had we not examined the zenplot for the strongest ACFs (as determined by the minimum p-value over the several Ljung-Box tests for each) over all components, this spurious auto-correlation might never have been discovered.

Recall that \texttt{TEL} is one of four components of the 465 fitted that
  had missing data at the beginning of their series, and that these were filled
  in by the first non-missing value (see Figure~\ref{fig:SP500:NA}).  One of
  these (\texttt{TWC}; Time Warner Cable) had very few missing cases; the two
  remaining (\texttt{DAL} for Delta Air Lines and \texttt{DFS} for Discover
  Financial Services) had about as many filled in as had \texttt{TEL}.  An
  examination of the standardized residuals for \texttt{DAL} and \texttt{DFS}
  showed a single enormous (several orders of magnitude) positive residual at
  the first observed (that is, non-filled) value in each series; there was no
  corresponding outstanding standardized residual estimated for \texttt{TWC}.
  Because each of \texttt{DAL} and \texttt{DFS} had only a single outlying
  estimated residual, neither component showed up in the strongest ACFs; the
  corresponding lag 1 plots resembled that of
  Figure~\ref{fig:SP500:zenplot:residual:check:LB:ACF_problem} except that no
  point appeared in the top right corner and hence no spurious
  auto-correlation. It would seem that the problem is an artefact of the fitting
  mechanism \texttt{ugarchfit()} from version 1.3-6 of the \texttt{rugarch}
  package together with the fact nearly the first $20\%$ of the series is
  constant (due to filling in missing values).  It is curious however, that
  \texttt{ugarchfit()} produced \emph{two consecutive} large residuals in the
  case of \texttt{TEL}, but only a single gigantic residual for each of
  \texttt{DAL} and \texttt{DFS}. A.~Ghalanos (maintainer of \texttt{rugarch})
  suggests that the problem lies in the number of constant values at the
  beginning of each series (personal communication).  Whatever the reason for
  this peculiarity in the fitting, it might easily have gone unnoticed had we
  not considered the zenplots of
  Figures~\ref{fig:SP500:zenplot:residual:check:LB}.

The zenplots of Figures~\ref{fig:SP500:zenplot:residual:check:LB} and
\ref{fig:SP500:zenplot:residual:check:LB:squared} show that the sixteen ACFs which most challenge the hypothesis of serial independence provide little evidence against the hypothesis of serial independence.  Other than the spurious lag one autocorrelations, the remaining autocorrelations are mostly quite small, especially for the standardized residuals of  Figure~\ref{fig:SP500:zenplot:residual:check:LB}.  For the volatility, a few series of Figure \ref{fig:SP500:zenplot:residual:check:LB:squared} show the occasional lag beyond the 95\% limits for uncorrelated data (given by the horizontal dashed lines).  On the basis of these plots,  a more complex $\GARCH$ component for these individual series might be considered, perhaps depending on the component involved, but care needs to be taken to not over interpret the significance of the observed departures.  They were, after all, selected as the most challenging cases (via 30 Ljung-Box tests) from $465$ components; the putative 5\% level given by the horizontal lines is a gross understatement of the actual observed level of significance.  Given the paucity of evidence against the hypothesis of serial independence, and since our primary modelling goal here is to
model the dependencies \emph{between} components,  we will continue our investigation using the common and parsimonious  $\ARMA(1,1)-\GARCH(1,1)$ for each of the 465 components. That is, we proceed based on our interpretation that these
``worst-case'' plots show little evidence against the null hypothesis of
serial independence in the estimated standardized residuals; more complex modelling of any remaining serial dependence structure will not be pursued.

\begin{remark}\label{rem:p:values:invalid:1}
  It is important to note that the plots of
  Figures~\ref{fig:SP500:zenplot:residual:check:LB} and
  \ref{fig:SP500:zenplot:residual:check:LB:squared} are constructed from
  zenpaths which, as with earlier plots, rely on some measure of each plot's
  importance to determine its order along any zenpath.  These particular figures
  used the minimum p-value from several Ljung--Box test statistic to provide the
  ordering.  Unlike the test statistic itself, the p-value has the
  advantage that all plots are then compared using a common scale.

  However, it is important to note that using the p-values to provide order is
  not the same as reporting them as the observed level of significance when testing the hypothesis of serial independence anywhere in the series. Because many tests are involved the actual significance level is necessarily
  larger than the smallest p-value, which was used here.
  After all, the p-value reported by each test assumes that it is the only test conducted.
  In contrast, when constructing the zenplots a
  great many tests
  were conducted and only the smallest p-values selected
  and reported (in this case from more than $13,000$ tests!). A true
  observed significance level must take into account this multiple testing.
  Determining
  how best to adjust the p-values so that they might accurately represent the
  true significance level is not always obvious, especially in this case.  For
  example, the adjustment here would need to account for the fact that first the
  minimum value is taken over 30 (not necessarily independent) tests, and then
  the minima of these minima is then taken over the 465 separate components.
  Adjusting the p-value to better reflect the true significance level is likely
  to be a challenge in this case, and in general for many other cases where
  plots are ordered.

  Fortunately, for the purpose of determining a zenpath ordering to be presented
  for display, the ordering provided by the (uncorrected) p-values is
  sufficient and will, in most cases, provide the same ordering as would the
  true observed significance levels.
  For this reason, wherever p-values are used to order plots it would be best to omit reporting the values so as not to mislead the viewer. \hfill $\qed$
\end{remark}
\subsection{Checking the marginal distributions for being $t$}\label{sec:model:check:margins:t}
We now turn to the model assumption of having $t$ innovations.  This assumption
can be assessed visually via a quantile-quantile plot (or Q-Q plot) of the
sample quantiles of the standardized residuals.

\begin{figure}[htbp]
  \centering
  \includegraphics[width=\textwidth]{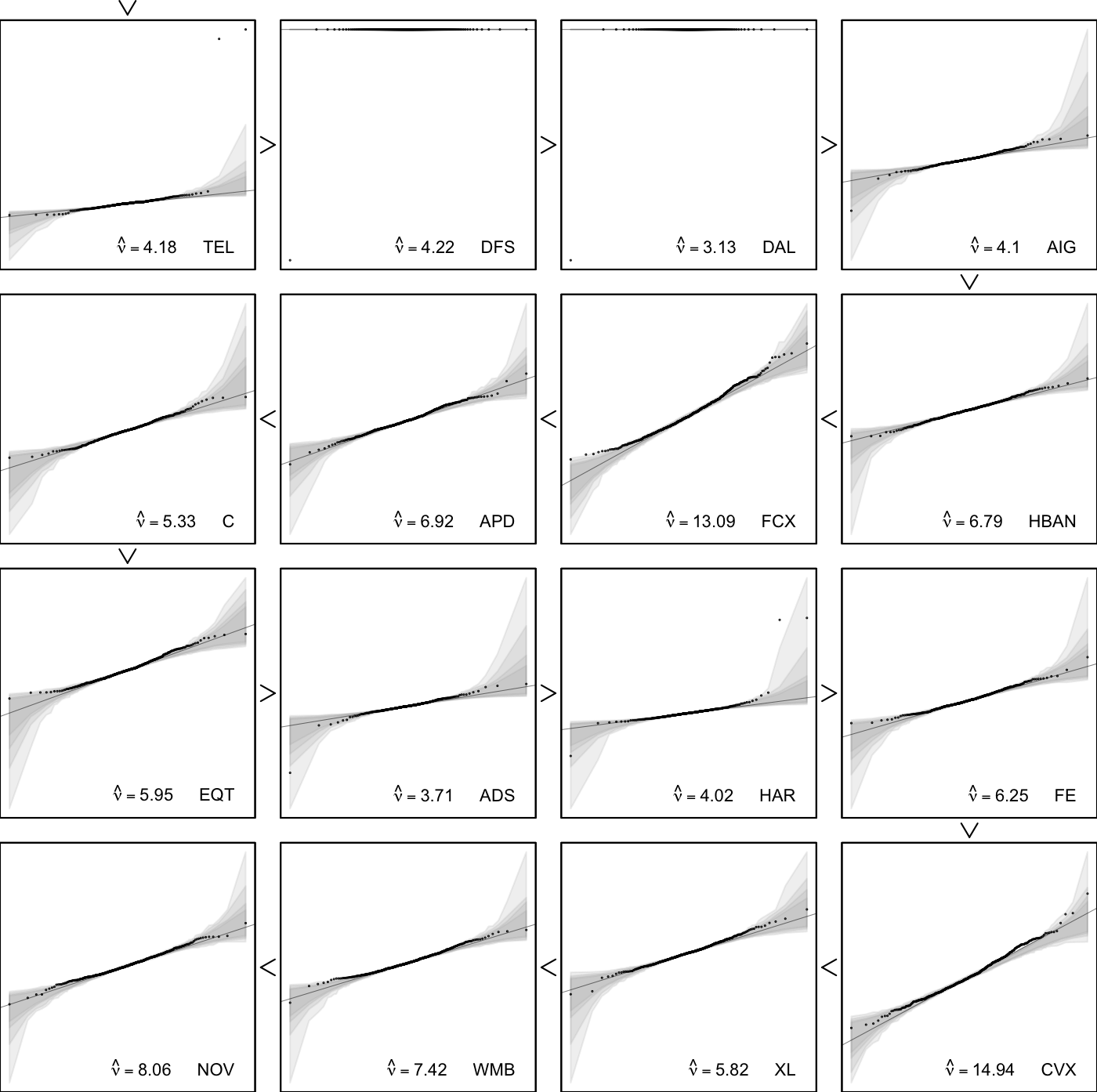}%
  \caption{A zenplot of Q-Q plots for those 16 margins having the largest
    Anderson--Darling test statistics; the fitted degrees of freedom
    ($\widehat{\nu}$) are as indicated in each
      plot. The line of exact agreement (the $y=x$ line) appears as a fine line
    in each plot.
  }
  \label{fig:SP500:zenplot:residual:check:AD}
\end{figure}

Figure~\ref{fig:SP500:zenplot:residual:check:AD}
shows Q-Q plots for sixteen components. The vertical axis of each Q-Q plot marks
the sample quantiles and the horizontal axis marks the theoretical quantiles for
a standardized $t$ distribution with degrees of freedom estimated from the
sample (displayed as $\widehat{\nu}$ in the plots). The grey regions are
point-wise empirical confidence envelopes for each quantile as constructed from
1,000 samples (each of size $n=755$) generated from a $t$ marginal on the
estimated degrees of freedom $\widehat{\nu}$.  From darkest to lightest, the
regions correspond to the central 90, 95, and 99\% of the simulated sample
quantiles as well as the range of all 1,000 generated.  Departures from the
marginal $t_{\widehat{\nu}}$ hypothesis are seen as points appearing at the
extremities (or outside) of the given grey regions; see \cite{oldford2016qqtest}
for details. The sixteen components chosen were those with largest
Anderson--Darling test statistic, that is, the zenpath is the path of
decreasing evidence against the null hypothesis that the marginal
$\ARMA(1,1)-\GARCH(1,1)$ residuals follow a standardized $t_{\widehat{\nu}}$
distribution with estimated degrees of freedom $\widehat{\nu}$.

The first three stocks \texttt{TEL} (TE Connectivity Ltd.), \texttt{DFS}
(Discover Financial Services), and \texttt{DAL} (Delta Air Lines) thus display
the strongest evidence against the null hypothesis. This is perhaps not too
surprising
in light of the discussion surrounding the ACF plots of Figures~\ref{fig:SP500:zenplot:residual:check:LB} and \ref{fig:SP500:zenplot:residual:check:LB:squared} as well as the residual plots of Figure~\ref{fig:SP500:zenplot:residual:check:LB:ACF_problem}. There we found that \texttt{TEL} had two large standardized residuals that caused a spurious lag 1 autocorrelation.  These now both appear again as the pair of large positive residuals in the first Q-Q plot of Figure~\ref{fig:SP500:zenplot:residual:check:AD}.   The next two Q-Q plots bring attention to the massive singleton outliers of \texttt{DFS} and \texttt{DAL} which, as reported earlier, are likely to be artefacts of the \texttt{ugarchfit()} fitting mechanism for series whose first values are all identical.  Again, by examining the zenplot of Figure~\ref{fig:SP500:zenplot:residual:check:AD} these difficulties are immediately brought to the attention of the analyst.

Similarly, it should be no surprise that \texttt{AIG} (American International
Group) shows up next in order as not having a $t$ innovation distribution.
During the subprime mortgage crisis of 2008 AIG stock collapsed and would have
failed entirely had it not been bailed out.  This is picked up
visually by the Q-Q plot as several of the positive standardized residuals are
outside the range of the simulated envelopes on the right where the envelope is
tight, just before it fans out.

It is interesting to note that the Anderson--Darling test statistics used to
order the Q-Q plots in Figure~\ref{fig:SP500:zenplot:residual:check:AD} change
abruptly in the supposed (uncorrected) p-values which they report. This is
the case, for example, after the first four plots appearing in the top row, at
\texttt{HBAN} (Huntington Bancshares) along the zenpath.

In contrast, several of the remaining Q-Q plots clearly show evidence against
the hypothesis of the $t$ marginal (with those estimated degrees of freedom
$\widehat{\nu}$) For example, \texttt{HAR} (Harman International Industries,
Inc.) and \texttt{ADS} (Alliance Data Systems, Inc.)  have outlying standardized
residuals, while \texttt{FCX} (Freeport-McMoRan), \texttt{C} (Citigroup, Inc.),
\texttt{EQT} (EQT Corp.), \texttt{CVX} (Chevron Corporation), and \texttt{WMB}
(Williams Companies, Inc.) each suggest some asymmetry.

Graphical methods are able to detect many different types of departures and,
with zenplots it actually becomes feasible to view at least the most interesting
ones among 465 Q-Q plots.  Some other measure based entirely on the geometric
features of the Q-Q plot might even be more helpful in choosing an interesting
zenpath than this Anderson--Darling test.

\begin{remark}
  In constructing the zenplot of Figure~\ref{fig:SP500:zenplot:residual:check:AD}, the value of the Anderson--Darling test-statistics were used to order the plots.  Generally, ordering by the p-value is preferred but is equivalent to ordering by the test statistic in this particular case.
  The reader is again cautioned that any such p-values would need to be corrected before being interpreted as observed levels of significance.  In addition to the problems of multiple testing mentioned
  in the earlier remark, to arrive at the correct level of significance,
  adjustment would also need to consider the effect of using degrees of
  freedom for the $t$ distributions that themselves are uncertain since they
  too must be estimated from the data.

  Again, the (uncorrected) p-values are used to determine an
  order in which to lay out the plots, beginning from those most likely to
  show evidence against the hypothesis (by this measure) to those least
  likely.  Perhaps, as the above analysis suggests, had we some measure of
  departure from the distributional hypothesis based only on the geometric
  configuration of a Q-Q plot, we might use that measure to order the plots --
  without any knowledge of what the corresponding level of significance might
  be.  The zenpath would serve up those plots which were of interest according
  to this measure. \hfill$\qed$
\end{remark}

\subsubsection{Zenplot layout}\label{sec:model:check:layout}
Some important features of zenplot construction enabled the compact layout of
Q-Q plots as in Figure~\ref{fig:SP500:zenplot:residual:check:AD} and make
practical viewing of all 465 plots possible.

First, a zenplot is actually a
layout of an alternating sequence of single coordinate and two-coordinate plots;
in Figure~\ref{fig:SP500:zenplot:residual:check:AD} the alternation is between a
``V'' arrow shape (one dimensional location indicating an order) and a Q-Q plot
(having two coordinates to be plotted).

Second, the location of each plot is
determined by where it appears in a sequence of ``direction'' indicators, one of
``u'', ``d'', ``l'', or ``r'', for ``up'', ``down'', ``left'' or ``right''; each
one is the directive for where the \emph{next} plot will appear relative to the
present plot.  The plots of Figure~\ref{fig:SP500:zenplot:residual:check:AD}
begin as the top left ``V'' drawn to indicate ``d'' for the relative position of
the first Q-Q plot.  It then follows a series of eight ``r'' directions to move
across the page alternating Q-Q plots and arrows, then two ``d'' to place the
arrow down and to position the following Q-Q plot down on the next row.  This
Q-Q plot exits left (direction ``l'') to start the sequence moving leftward back
across the page.  The zenplot thus zigzags back and forth across the page moving
down when the edge of the page is reached and then reversing the horizontal
direction. The zenplots of all previous figures were constructed using a
slightly more complex sequence of directions; see the demo \texttt{SP500} for
the corresponding source code.  By using an appropriate sequence of directions,
essentially any layout pattern of alternating plots can be constructed with a
zenplot (for example, a spiral; see \texttt{?zenplot} for more details and
additional features not described here).

Third, there is no restriction on the
plots that may be drawn.  The Q-Q plot used in
Figure~\ref{fig:SP500:zenplot:residual:check:AD}, for example, is not from the
\texttt{zenplots} package of \cite{zenplots} but from \texttt{qqtest} of
\cite{oldford2016qqtestPackage}.  If a plot can be drawn in any one of
three \R\ plotting systems of \texttt{graphics}, \texttt{grid} (including \texttt{ggplot2}), or
\texttt{loon}, it can be part of a zenplot.

\subsection{Comparing models}\label{sec:full:t} %
In modelling high dimensional data, we have seen how zenplots are useful for
both model interpretation and for model assessment.  When we have different but
competing models available, more direct comparisons of the fitted models can
lead to deeper insight both about the dependence characteristics of the data and
about the relative merits of the models.  Again, zenpaths and zenplots can be
put to good use in this sort of analysis for high-dimensional data.

To be concrete, the approach taken so far has been very flexible, in that it
allows separate $t$ copula models for every pair of variates (an even more
flexible nonparametric approach is sketched in Appendix
\ref{sec:taildep:biv:nonparam}). This is sensible if we are only interested in
the notion of bivariate tail dependence.  It might be argued, however, that
there should be a single multivariate model fitted, one that is designed to
incorporate all 465 dimensions simultaneously.

One natural candidate to compete with a collection of pairwise $t$ copula models
would be a multivariate $t$ copula.  It is also known to have fit multivariate
financial return data fairly well and can actually be fit to our
high-dimensional data.  To fit the multivariate $t$ copula model we use the
fitting procedure of \cite{mashalzeevi2002}; see more recent versions of the \R\
package \texttt{copula} of \cite{copula}.

In Section~\ref{sec:taildep:t}, tail dependencies of the fitted multivariate $t$
model are compared to those of the previous collection of pairwise fitted $t$
copulas. Zenpaths derived from comparisons between these two fitted types of
models identify important differences in the tail-dependence estimates; the
corresponding zenplot presents the component pairs for further examination.  In
Section~\ref{sec:gof:full}, we use differences in the pairwise fits of these
models as measured by another Anderson--Darling test statistic to identify pairs
where shared model assumptions might be suspect.  The possibility of other
measures of interest is raised in Section~\ref{sec:gof:other}.

\subsubsection{Tail dependence}\label{sec:taildep:t}
As with the pairwise modelling, this fully joint model will have correlation
parameters from whose estimates, via Equation \eqref{eq:lamFromRho}, an estimate
of the upper tail dependence for any pair of returns can be obtained. These measures
from the full model can then be examined as before.

\begin{figure}[htbp]
  \centering
  \includegraphics[width=0.45\textwidth]{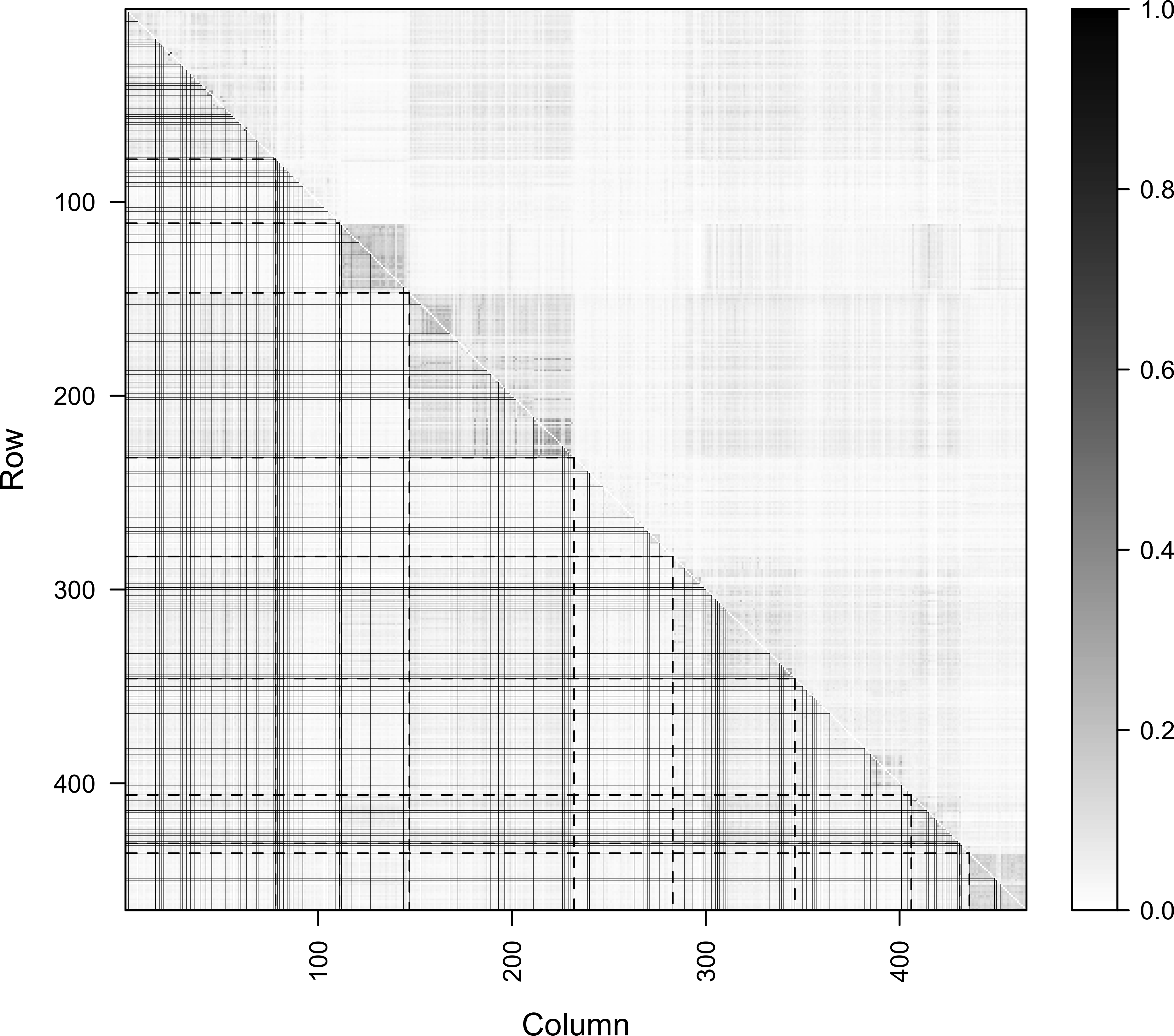}%
  \hfill
  \raisebox{4mm}{\includegraphics[width=0.51\textwidth]{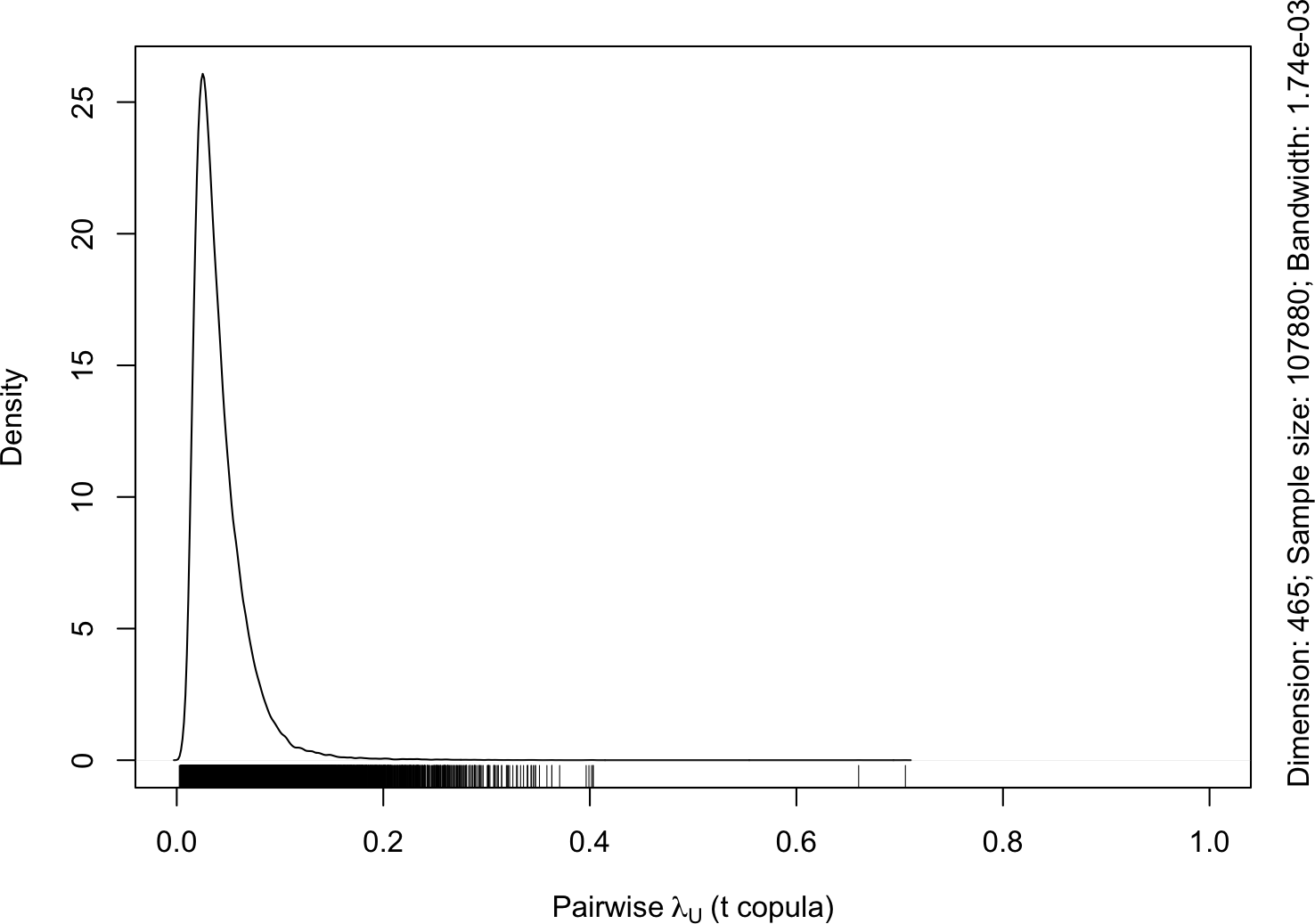}}%
  \caption{Matrix of implied upper tail-dependence coefficients as
    obtained from a fitted joint $t$ copula (left-hand side); the fitted degrees
    of freedom were 12.98. Density plot of the corresponding
    $\binom{d}{2}$ entries (right-hand side).}
  \label{fig:SP500:Lambda:joint:t}
\end{figure}

Figure~\ref{fig:SP500:Lambda:joint:t},
is the equivalent of Figure~\ref{fig:SP500:Lambda}, except now the
$\widehat{\lambda}$s have been estimated from the \emph{full}
465-dimensional, $t$ copula model. The matrix display at left is similar in
pattern to its counterpart from Figure~\ref{fig:SP500:Lambda}.  The most
dramatic difference between the two is that the matrix of
Figure~\ref{fig:SP500:Lambda:joint:t} looks like a washed out version of that of
Figure~\ref{fig:SP500:Lambda}, that is, the full model overall suggests weaker pairwise upper
tail dependencies than do the separate pairwise models.  Comparing the two densities
of the two sets of estimates from the right hand displays of
Figures~\ref{fig:SP500:Lambda} and \ref{fig:SP500:Lambda:joint:t}, we see that
the full model has reduced the range of the bulk of the dependence estimates
from less than 0.3, to less than 0.1.

\begin{figure}[htbp]
  \centering
  \begin{tabular}{ccc}
    \includegraphics[height=0.22\textheight]{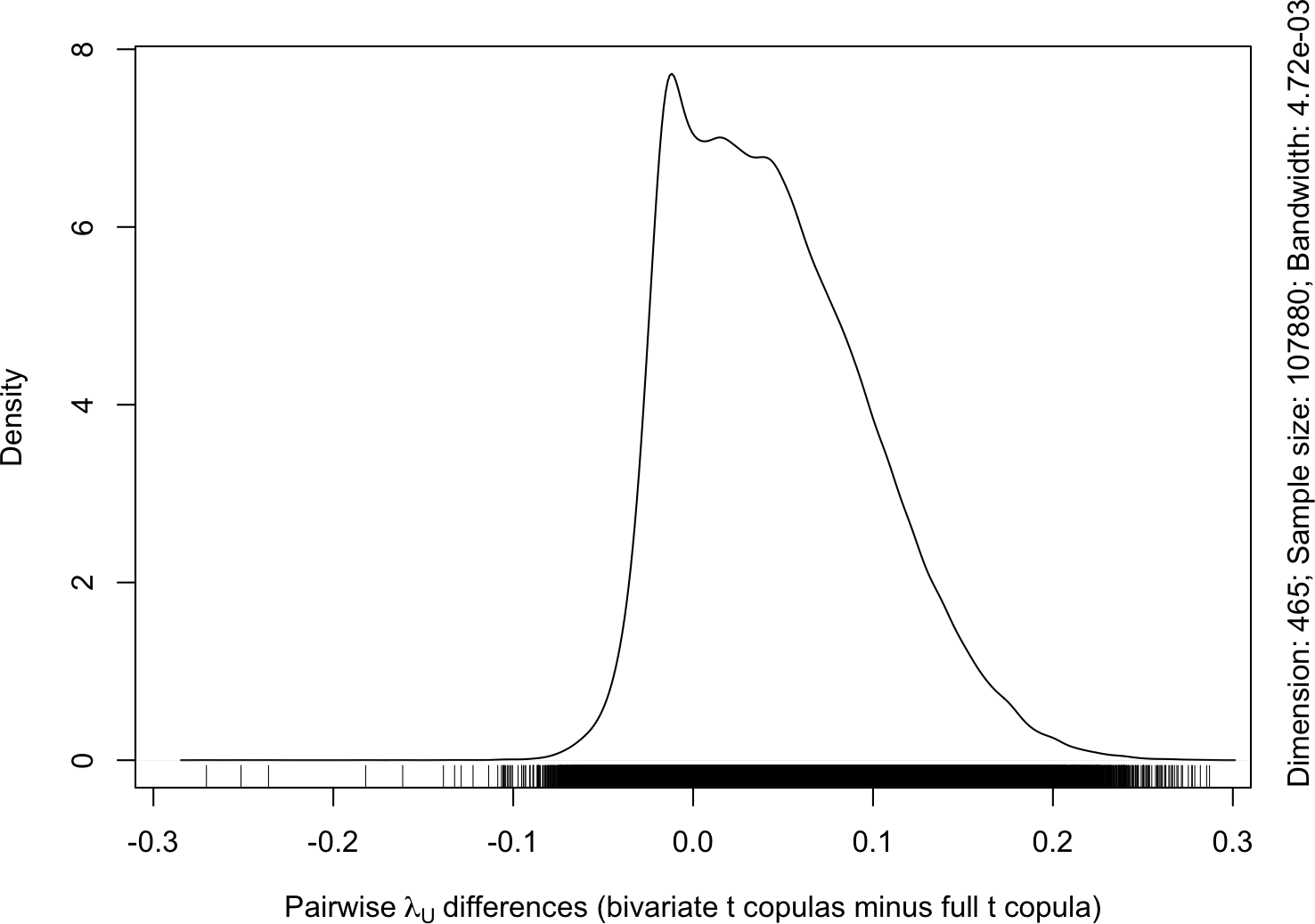}%
   & & %
  \includegraphics[height=0.22\textheight]{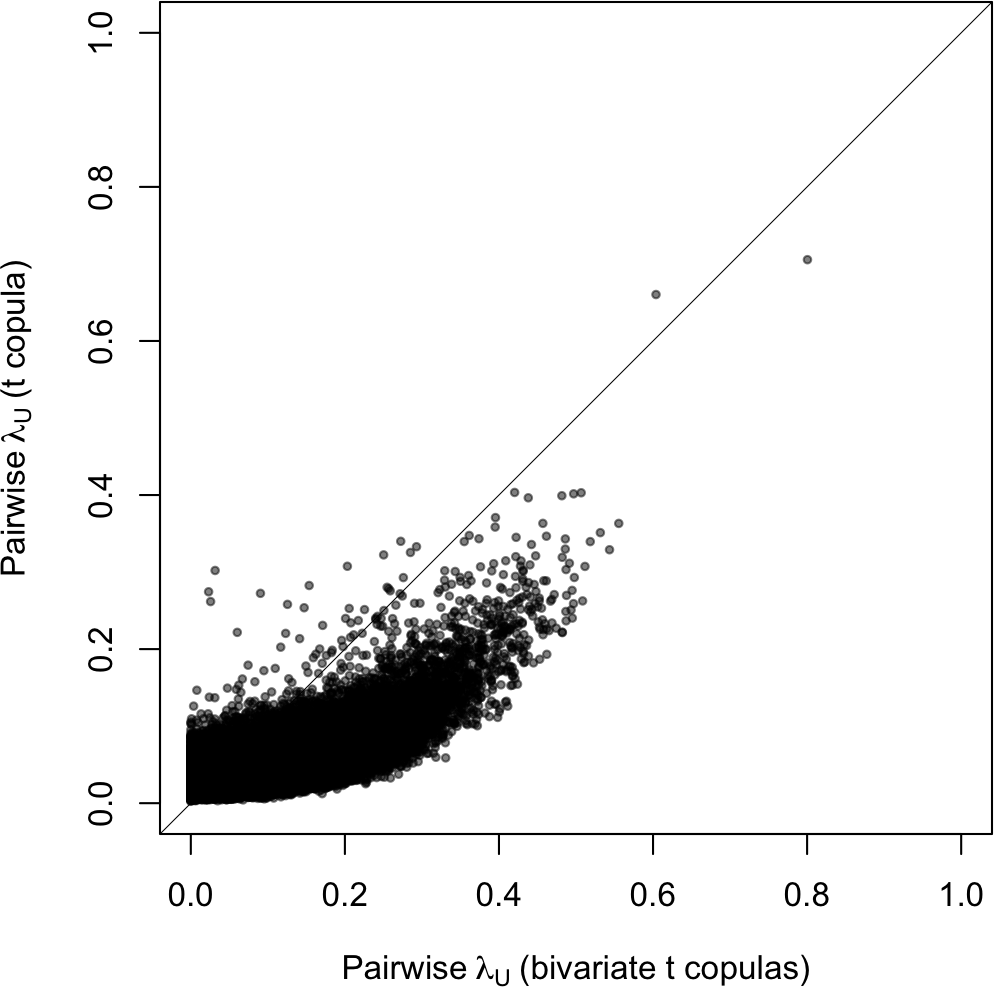}%
   \\
  \footnotesize{(a) Differences between $\widehat{\lambda}$.}
  & &
  \footnotesize{(b) Coefficients $\widehat{\lambda}$.}  \\
  &&\\
  \includegraphics[height=0.22\textheight]{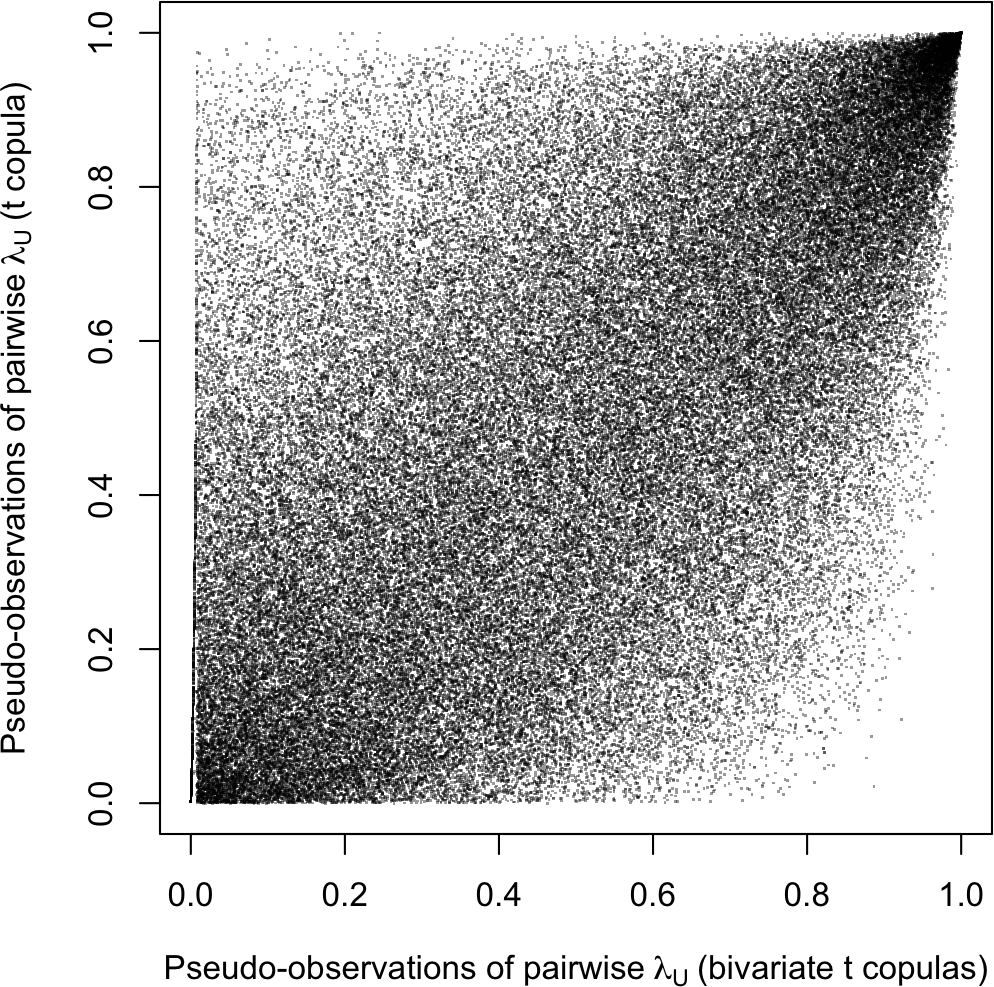}%
  & & %
  \includegraphics[height=0.22\textheight]{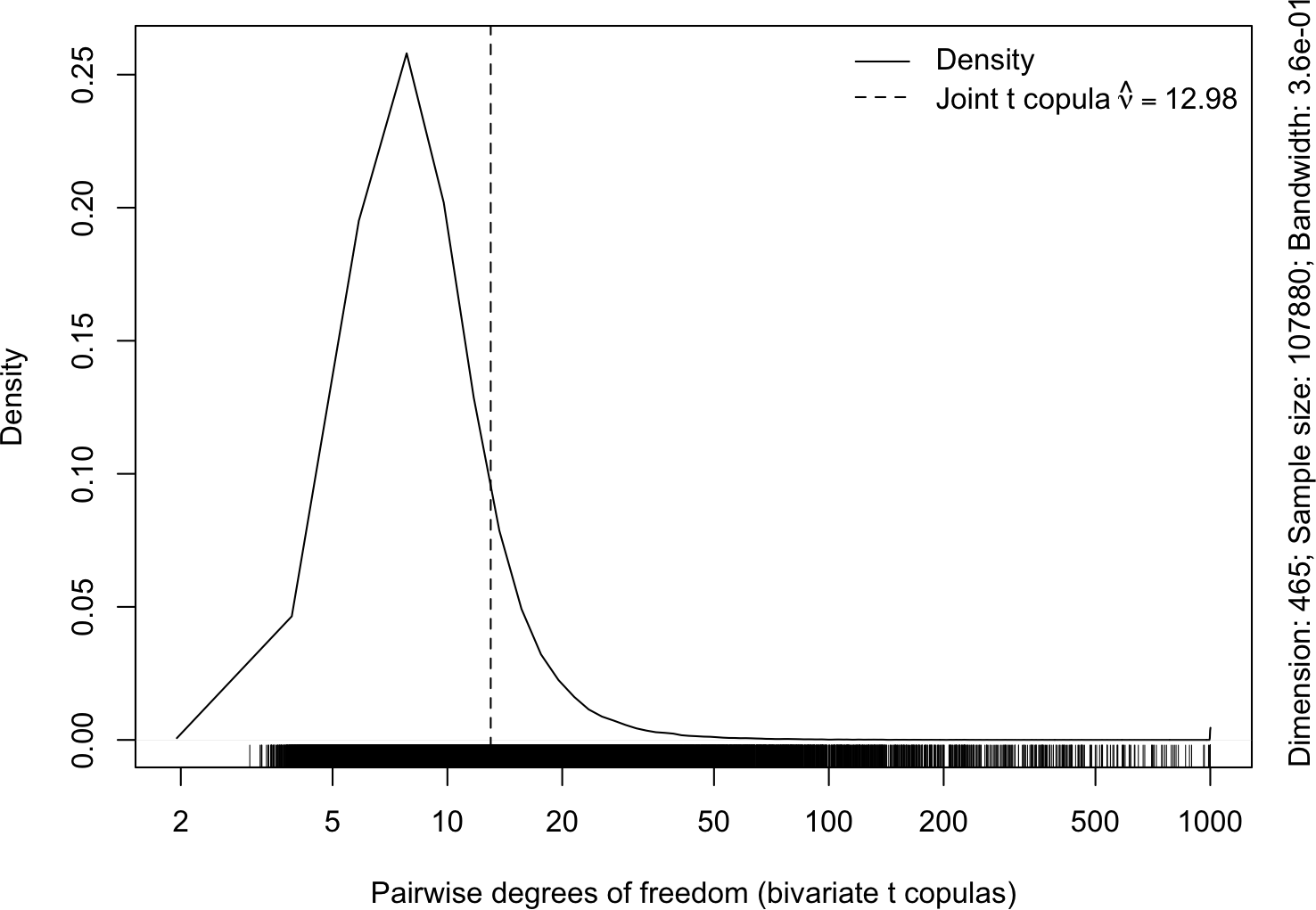}%
\\
  \footnotesize{(c) Pseudo-observations of the $\widehat{\lambda}$s.}
  & &
  \footnotesize{(d) Degrees of freedom.}
  \end{tabular}
  \caption{Comparing estimates: joint $t$ copula and the separate pairwise $t$
    copula models.}
  \label{fig:SP500:joint:t:pairwise:compare}
\end{figure}

Matching corresponding estimates of $\lambda$ from the two models, we can
compare their values more directly.
Figure~\ref{fig:SP500:joint:t:pairwise:compare}(a)
displays the density of their differences showing that the estimates from the
separate pairwise models are typically larger (often by more than $0.1$) than
those from the joint model.  Figure~\ref{fig:SP500:joint:t:pairwise:compare}(b)
plots these estimates as pairs; the points lying below the $y=x$ line are cases
where separate pairwise modelling estimates larger upper tail dependencies than
does the joint copula model. Dependence between the two sets of upper
tail-dependence coefficient estimates is summarized by the pseudo-observations
from the upper tail-dependence coefficients plotted in
Figure~\ref{fig:SP500:joint:t:pairwise:compare}(c) which suggest an asymmetric
dependence.

\begin{figure}[htbp]
  \centering
  \includegraphics[width=\textwidth]{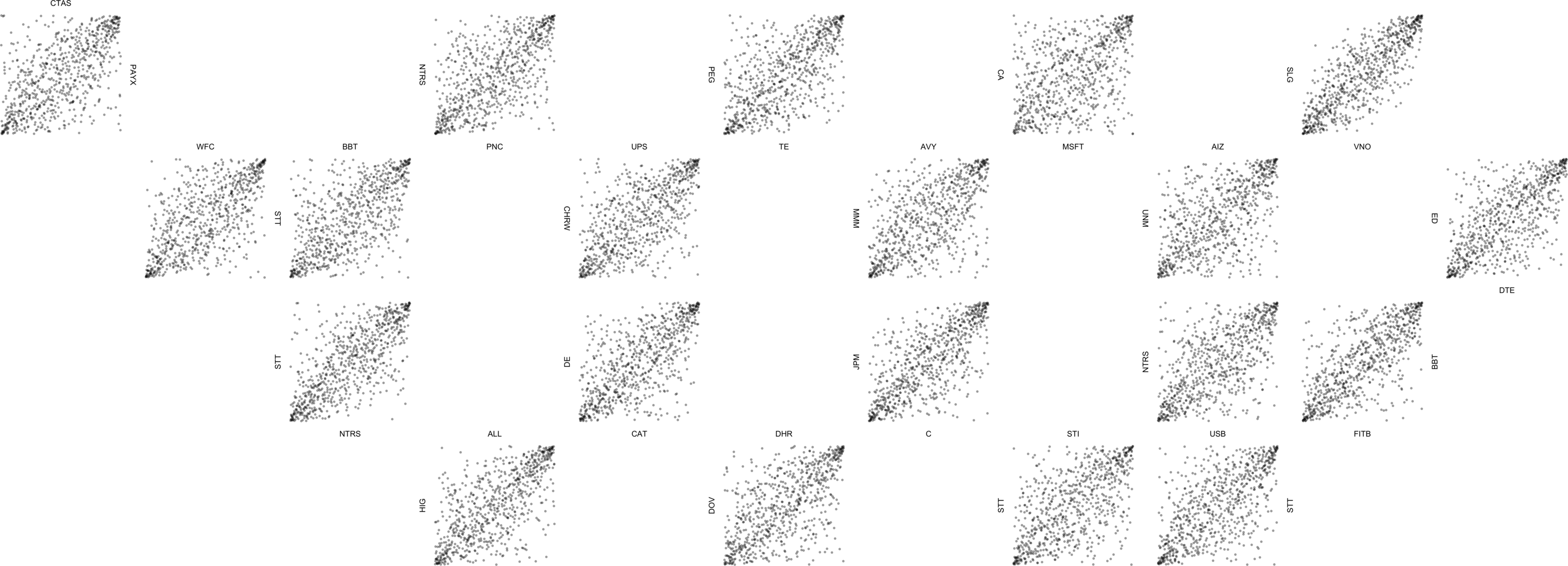}%
  \caption{A zenplot of 20 component pairs whose difference in upper tail-dependence estimates are greatest between a pairwise $t$ copula and a 465 dimensional multivariate $t$ copula.  Components are sometimes from the same sector, sometimes from different sectors.}
  \label{fig:SP500:Lambda:modeldiff:t}
\end{figure}

With different estimates of the upper tail-dependence coefficients arising from
the two fitted models, it might be of interest to examine which pairs of
components have given rise to the greatest differences.
Figure~\ref{fig:SP500:Lambda:modeldiff:t}
shows the zenplot of the twenty component pairs having the greatest absolute difference in upper tail-dependence coefficients.

We might also investigate the difference in estimated degrees of freedom from the two models.
The density of the degrees
of freedom from the pairwise models is shown in the right-hand plot of
Figure~\ref{fig:SP500:joint:t:pairwise:compare}; the single estimated degrees of
freedom of the full $t$ copula, 12.98, is shown as the vertical line.  As can be
seen, the pairwise models suggest heavier tails (lower degrees of freedom) in
general than does the joint model.  The dramatic restriction on the degrees of
freedom forced by the full model seems to force lighter tails on most of these
pairwise $t$ copulas.

Unconstrained, the degrees of freedom from the pairwise fits are highly variable
and, for the most part, lower than that estimated by the full model.  As with
the estimated coefficients of upper tail dependence, a zenplot could be
constructed for those pairs of components where the difference in the estimated
degrees of freedom is largest, or, given that tail dependence is of primary
interest, perhaps just those pairs having the smallest estimated degrees of
freedom in the pairwise models.  The zenplot of the latter shares many of the
same pairs as those shown in Figure~\ref{fig:SP500:Lambda:modeldiff:t} and so is
not produced here.

Instead, in the next subsection, we introduce another measure based on
both types of models which will be used to produce another zenplot of
interesting component pairs.

\subsubsection{Comparing fits}\label{sec:gof:full}

Suppose we had a test that assessed the fit of a copula model on any pair
of components.  For every pair of components we could then test each of our
models: the pairwise fitted $t$ copula models and the corresponding marginal
models of the full multivariate $t$ copula model.  If one doesn't fit but the
other does, then this suggests that the degrees of freedom and correlational
structure of one model better describes this pair of components than does the
other; there would seem to be no evidence against the distributional shape in
this case.  However, if \emph{neither} fit well then it might reasonably be
asserted that the distributional family is sufficiently suspect that a plot of
the pseudo-observations for this pair should be examined.  A zenplot of all
component pairs where both types of models fit poorly might therefore be of
interest to an analyst.

Such a test can be constructed as an Anderson--Darling test based on observations derived from (a function of) the
transformation of \cite{rosenblatt1952} of the pseudo-observations common to both models.
In each case, the transformation will be derived from the hypothesized copula model.
The derived observations are constructed as follows.  For a random vector
$\bm{U}\sim C$ for some copula $C$, we can transform
$\bm{U} = (U_1, \dots, U_d)$ to $\bm{V}$ as
\begin{align*}
  V_1&=U_1,\\
  V_2&=C_{2|1}(U_2\,|\,U_1),\\
      &\phantom{:}\vdots\\
  V_d&=C_{d|1,\dots,d-1}(U_d\,|\,U_1,\dots,U_{d-1}),
\end{align*}
where, for $j\in\{2,\dots,d\}$, $C_{j|1,\dots,j-1}(u_j\,|\,u_1,\dots,u_{j-1})$
denotes the conditional probability that $U_j\le u_j$ given
$U_1=u_1,\dots,U_{j-1}=u_{j-1}$. This embeds the conjectured copula in
the calculations and, following \cite{rosenblatt1952}, the transformed random
vector $\bm{V}$ will follow a $\U(0,1)^d$ distribution if and only if the copula $C$ is
correct; note that the construction depends on the order of the variates in the
Rosenblatt transformation.
Any mapping $g: \left[0, 1 \right]^d \rightarrow \Reals$ will produce a random
variable $W = g(\bm{V})$ whose distribution, if known, may be tested via
an Anderson--Darling test.

For illustrative purposes, we choose
$W=\sum_{j=1}^d\left(\Phi^{-1}(V_{j})\right)^2$ where $\Phi$ is the standard
normal distribution function. If $C$ is correct, $W \sim \chi^2_d$ and so could
be assessed using the realizations $w_1, \ldots, w_T$. Unfortunately, the
realizations $w_1, \ldots, w_T$ are not directly available and the estimated
realizations $\widehat{w}_1, \ldots, \widehat{w}_T$ must be used in their place.
Clearly the latter will not be independent $\chi^2_d$ realizations though they
may be good enough for this purpose (as suggested, for $d=2$, by
\cite{breymanndiasembrechts2003}; the limitations of this tests are known well,
see, for example, \cite{dobricschmid2007}).  When interest lies only in pairs of
components, then we could use each fitted copula model as the hypothesized
copula for the Rosenblatt transformations on each pair to determine putative
$\chi^2_2$ (estimated) realizations, and hence an Anderson--Darling test, for
each copula on every pair.

Anderson--Darling tests were constructed in this way and applied to all
  $107,880$ pairs of components, once using the fitted pairwise $t$ copulas and
  once using that corresponding bivariate margin of the full multivariate $t$
  copula. For pairs with small p-values in both cases, the choice of the $t$
  family may be a problem (and not just the common degrees of freedom assumed by
  the full model, for example). For pairs of returns with rather small p-values
  when testing based on the full model but rather large p-values for the
  pairwise model, a bivariate $t$ copula may be an adequate model for each of
  these pairs but not a full $t$ copula model.

\begin{figure}[htbp]
  \centering
  \includegraphics[width=0.95\textwidth]{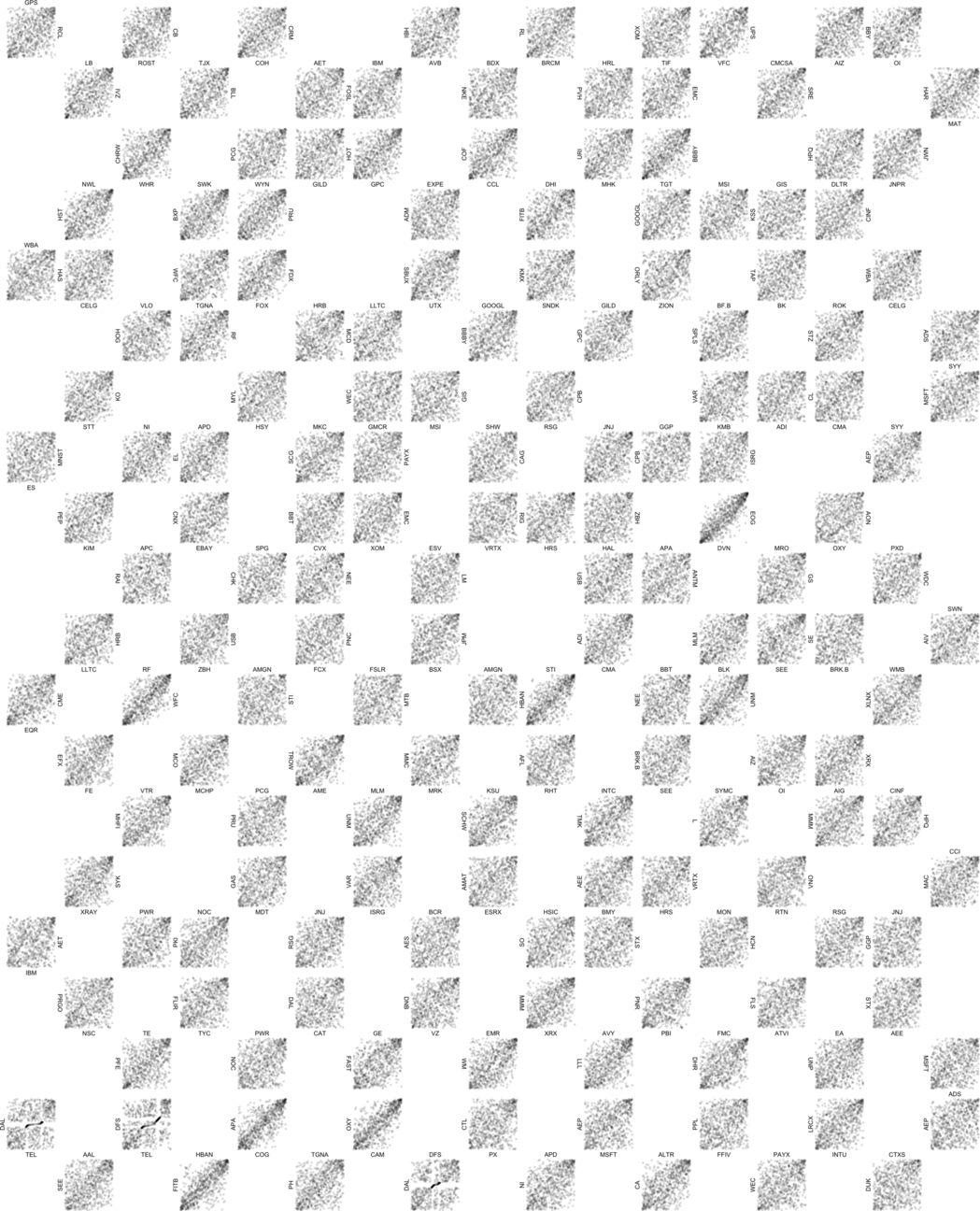}%
  \caption{A zenplot constructed from a zenpath displaying the
    pseudo-observations of those pairs of variables with smallest p-value of
    the Anderson--Darling test for the bivariate $t$ copulas. Note that all
    pairs until and including \texttt{DFS} and \texttt{DAL} (see last row) are
    precisely those for which the order is the same as for the p-values for the
    bivariate $t$ copulas implied by the full $t$ copula model. %
  }
  \label{fig:SP500:zenplot:pobs:biv:t:copulas:smallest:p:value}
\end{figure}

Figure~\ref{fig:SP500:zenplot:pobs:biv:t:copulas:smallest:p:value}
lays out the pseudo-observations of those pairs of standardized residuals with
smallest p-values for the Anderson--Darling test for the bivariate $t$ copula
models. Note that all pairs until and including \texttt{DFS} and \texttt{DAL} in
the last row of the zenplot are precisely those for which the order is the same
as for the p-values for the bivariate $t$ copulas implied by the full $t$ copula
model.  Furthermore, whenever adjacent plots share variates they are displayed
in the same group (as determined by \texttt{connect\_pairs(, duplicate.rm =
  TRUE)}). The pseudo-observations may now be critically examined by an
experienced analyst to determine how each pair of standardized residuals appears
to depart from a $t$ copula and to possibly even suggest an alternative copula
family. Note that this is not an easy task as replacing a bivariate model by
another one can (and most often will) not lead to a proper multivariate model
anymore; however, the information on \emph{how} the data departs from the
hypothesized model is valuable in making a decision whether the latter is still
acceptable for the modelling task at hand.

A visual inspection can reveal departures other than those identified by the
test statistic used to select the plot.  For example, Anderson--Darling
emphasizes departures in the hypothesized distribution's tails.  The
hypothesized distribution here, however, is $\chi^2_2$ (albeit derived from a
hypothesized $t$ copula).  The right tail of this $\chi^2_2$ is constructed from
large \emph{and} small pseudo-observations and so does not necessarily identify
a difference between the upper-right and lower-left of the plots of
pseudo-observations.  Rather, these two tails of the bivariate $t$ copula are
treated together, as one, in the assessment. Even so, examination of the plots
of Figure~\ref{fig:SP500:zenplot:pobs:biv:t:copulas:smallest:p:value}
does reveal several showing an apparent asymmetry between the two densest
corners, a well-known stylized fact of financial data not captured by a $t$ copula.
Asymmetry in the other direction can also be detected visually in several plots.

Similarly, the left tail of this $\chi^2_2$ is constructed from
pseudo-observations near $1/2$ and so this Anderson--Darling test will
emphasize differences from the centre of a $t$ copula.  Unusual spaces and/or
densities in the centre of the plots might therefore be detected by this test.
Indeed, the last plot (bottom-leftmost plot) of
Figure~\ref{fig:SP500:zenplot:pobs:biv:t:copulas:smallest:p:value}
shows an extremely unusual central configuration.  The returns involved are
those of \texttt{DAL} and \texttt{DFS} which were also previously identified by
the Q-Q plots of Figure~\ref{fig:SP500:zenplot:residual:check:AD}.  These were two
of the three stocks that were missing many measurements early in the time period
under study.  This would seem to corroborate the value of the Anderson--Darling
test on the middle part of the copula.  Unfortunately, when the third incomplete
stock, \texttt{TEL}, is plotted against either of \texttt{DAL} or \texttt{DFS},
the pseudo-observations also reveal strong linear (non $t$ copula) patterns at
the centre.  Yet, neither plot is highlighted by Anderson--Darling.

Another problem with our use of the Anderson--Darling test revealed is
numerical. Of the plots of
Figure~\ref{fig:SP500:zenplot:pobs:biv:t:copulas:smallest:p:value},
all until that of \texttt{DAL} and \texttt{DFS} have the same
calculated p-value.  This is because the squared normal probability
integral transform for some of the putative uniforms actually returned
\texttt{Inf} in \R. This may not be such a problem for purposes of identifying
pairs that are not $t$ copulas, but it severely diminishes the utility of an
ordered layout provided by zenplot. Those plots of
Figure~\ref{fig:SP500:zenplot:pobs:biv:t:copulas:smallest:p:value}
are thus unordered with respect to one another (though they do properly precede the
last one).

Other criticisms can and have been levelled at the Anderson--Darling test and,
as noted earlier, it would be better
were it replaced
by a parametric bootstrap; see \cite{dobricschmid2007} and
\cite{genestremillardbeaudoin2009}). Besides the challenge
of determining arbitrary values of the cumulative distribution function of a
multivariate $t$, with non-integer degrees of freedom, the computational burden
alone of a parametric bootstrap for all pairs of such high-dimensional data is
currently prohibitive for practical purposes.  In our problem, we have $d=465$
yielding $107,880$ bivariate pairs of $n=755$ pseudo-observations.  By our
calculations, the computation required to determine p-values in this
way, for this problem, is possibly only half the computation needed by
\cite{genestremillardbeaudoin2009} to carry out their entire Monte Carlo
experiment restricted to $d=2$.  And that, they report in 2009, ``required the
nearly exclusive use of 140 CPUs over a one-month period.''

\subsubsection{Other measures}\label{sec:gof:other}
It is an important feature of a zenplot that all pairs may be filtered by some
measure of interest so that only those pairs which matter are presented to the
analyst for closer examination.  And to this end, the Anderson--Darling test
employed above  can be of value, but can also be problematic as was just discussed.
However, it would be better to have several such
measures of interest, each detecting a different aspect of the data.

A zenplot
could then be produced for every such measure and provide the analyst much more
insight into the data.  For general characteristics of a scatterplot,
scatterplot diagnostics (or \emph{scagnostics}) were introduced about 30 years ago
by John and Paul Tukey and implemented more recently by
\cite{wilkinson2005graph}.  It is an interesting research problem to define
analogous measures of interest but which are specific to copula modelling.

Even without such measures, for dimensions as high as $d=465$ examined here,
zenplots make it possible to actually examine all $d \choose 2$ plots. The
compactness of a zenplot is such that as many as 660 plots can be produced per
page, giving about 164 pages of plots that serve at least as a visual record of
the dependencies between returns.  As proof of concept, we actually examined
$107,880$ plots using a standard PDF reader on screen in about 30 minutes.  Even
in this short of time, we were able to get a sense of the variety of patterns in
the pairs and to visually identify the most unusual pairs (for example, any two
of \texttt{DAL}, \texttt{DFS}, and \texttt{TEL}).

\section{Conclusion and discussion}\label{sec:con}
A zenplot is a zigzag-like structure which consists of pairwise (alternating
one- and two-dimensional) plots. The typical (but not exclusive) use-case is
where two consecutive pairwise plots share an axis, that is, a variate.  A
zenpath allows one to construct, in various ways, a path along pairs of
variates which can then be laid out with a zenplot.  One particularly important way
to construct a zenpath is to have measures of interest on each plot; then only those plots
that are interesting can be presented in the zenplot.

Zenplots and zenpaths are useful, for example, for detecting and visualizing
dependence in high-dimensional data when scatterplot matrices are too crowded,
not meaningful, or when it is computationally infeasible to display all
bivariate margins. We demonstrated such a case with \SP\ constituent data from
2007 to 2009, thereby focusing on the notion of pairwise tail
dependence.

Zenplots can equally well be used as graphical goodness of fit tools for
detecting regions or dimensions of departure from the assumed model.  In Section
\ref{sec:model:check}, a number of model assumptions were assessed graphically
by examining the strongest departures as given by a variety of common measures.

We also compared an ensemble of pairwise models against a single
high-dimensional model.  For example, the zenplots composed of Q-Q plots allowed
the marginal $t$ distributions implied by both types of models to be assessed as
well as whether the single degrees of freedom parameter (and thus a joint
multivariate $t$ model) was supported by the data.  Wherever model selection
methods lend themselves to a series of graphical displays, zenplots and zenpaths
will permit the displays to be efficiently laid out in an ordered fashion.

Finally, we saw that zenplots can be customized in quite a flexible way. Besides
providing one's own zigzag-like structure, groupings, and plotting functions,
one can also provide another graphics systems (available are \texttt{graphics}
(used here in this work), \texttt{grid} (and thus also \texttt{ggplot2}) and
even \texttt{loon} plots (for dynamic interaction)); see the many examples on
\texttt{?zenplot} for more details concerning these features.

\subsection*{Acknowledgments}
We would like to thank the associate editor for detailed feedback and many
helpful remarks. We would also like to acknowledge the support of NSERC
through our Discovery grants.

\appendix

\section{Tail dependence based on bivariate nonparametric estimators}\label{sec:taildep:biv:nonparam}
\begin{figure}[htbp]
  \centering
  \includegraphics[width=0.45\textwidth]{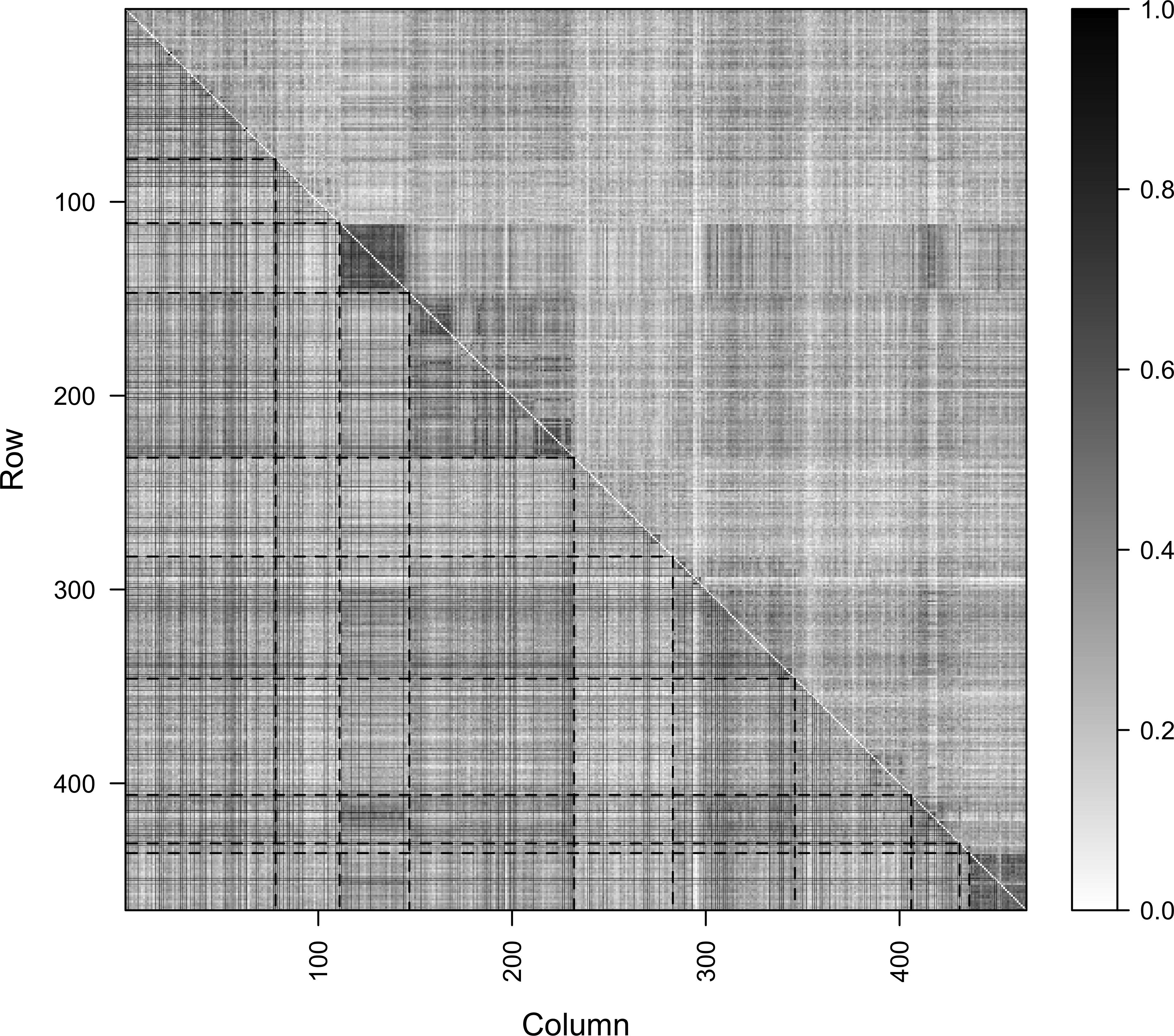}%
  \hfill
  \raisebox{4mm}{\includegraphics[width=0.51\textwidth]{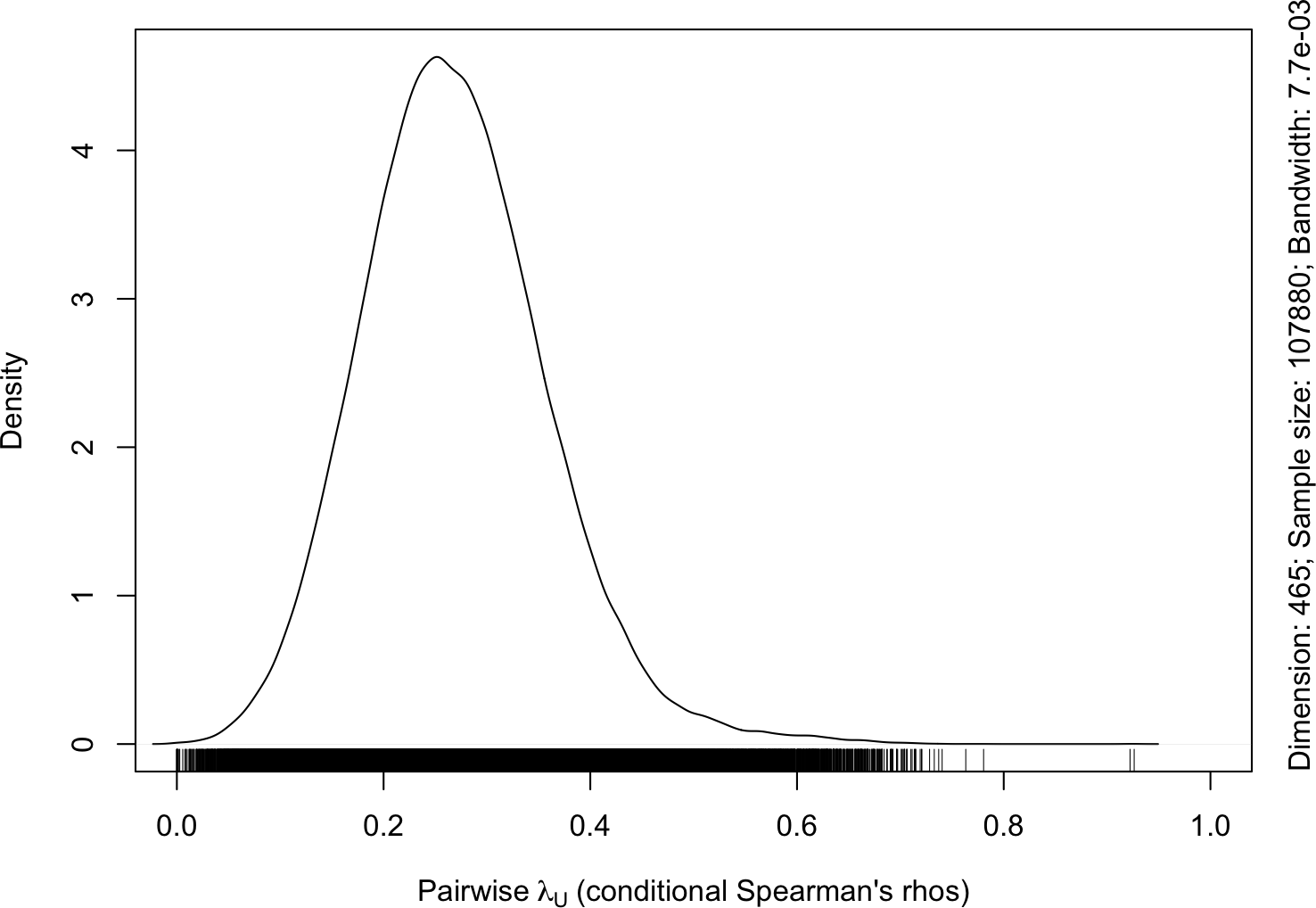}}%
  \caption{Matrix of pairwise upper tail-dependence coefficients as estimated nonparametrically by
    conditional Spearman's rho (left-hand side) and density plot of the corresponding
    $\binom{d}{2}$ entries (right-hand side).}
  \label{fig:SP500:Lambda:nonparam}
\end{figure}

Another estimator for upper tail dependence was suggested by
\cite{schmidschmidt2007}; see also
\cite[p.~231]{jaworskidurantehaerdlerychlik2010}. It is nonparametric and
essentially a properly scaled conditional Spearman's rho computed from the top
right corner of the bivariate
distribution. Figure~\ref{fig:SP500:Lambda:nonparam} shows the analogue plots to
Figures~\ref{fig:SP500:Lambda} and \ref{fig:SP500:Lambda:joint:t} for this
nonparametric estimator, were the top right corner is determined by the marginal
90\% quantiles, that is, those pseudo-observations which fall into $[0.9,1]^2$.

\printbibliography[heading=bibintoc]

\end{document}

%
%
%
%
